\def\figsize{9.5cm}
\def\figsiz{8.5cm}
\def\smtopskip{-1.2cm}
\def\smbotskip{-1.0cm}

\def\rn{\bibitem{}}
\def\nn#1 #2{#1, #2.}				
\def\nnn#1 #2 #3{#1, #2. #3.}			
\def\nnnn#1 #2 #3 #4{#1, #2. #3. #4.}		
\def\nnnnn#1 #2 #3 #4 #5{#1, #2. #3. #4. #5.}	
\def\dualand{, \&\hbox{ }}				
\def\multiand{, \&\hbox{ }}				

\def\rg#1;#2;#3;#4;#5;#6 {\par\rn#1 #2, {\it #3}, {\bf #4}, #5 (``#6'') \par}
\def\rf#1;#2;#3;#4;#5 {\par\rn#1 #2, {\it #3}, {\bf #4}, #5\par}
\def\rfbook#1;#2;#3;#4;#5 {{\frenchspacing\par\rn#1 #2, {\it #3} (#4: #5)\par}}
\def\rfproc#1;#2;#3;#4;#5;#6 {{\frenchspacing\par\rn#1 #2, in {\it #3}, ed. #4 (#5: #6)\par}}
\def\rfprocp#1;#2;#3;#4;#5;#6;#7 {{\frenchspacing\par\rn#1 #2, in {\it #3}, ed. #4 (#5: #6), p#7\par}}
\def\rfprep#1;#2;#3  {{\par\rn#1 #2, #3\par}}
\def\rfprepp#1;#2;#3 {{\par\rn#1 #2, #3\par}}

\def\Mpc{{\rm Mpc}}
\def\hperMpc{\,h/\Mpc}
\def\hMpc{\,h^{-1}\Mpc}

\def\expec#1{\langle#1\rangle}

\def\etal{{\frenchspacing\it et al.}}
\def\ie{{\frenchspacing\it i.e.}}
\def\eg{{\frenchspacing\it e.g.}}
\def\etc{{\frenchspacing\it etc.}}

\def\beq#1{\begin{equation}\label{#1}}
\def\eeq{\end{equation}}
\def\beqa#1{\begin{eqnarray}\label{#1}}
\def\eeqa{\end{eqnarray}}
\def\eq#1{equation~(\ref{#1})}
\def\Eq#1{Equation~(\ref{#1})}
\def\eqn#1{~(\ref{#1})}

\def\fig#1{Figure~\ref{#1}}
\def\Fig#1{Figure~\ref{#1}}

\def\sec#1{Section~\ref{#1}}

\def\spose#1{\hbox to 0pt{#1\hss}}
\def\simlt{\mathrel{\spose{\lower 3pt\hbox{$\mathchar"218$}}
     \raise 2.0pt\hbox{$\mathchar"13C$}}}
\def\simgt{\mathrel{\spose{\lower 3pt\hbox{$\mathchar"218$}}
     \raise 2.0pt\hbox{$\mathchar"13E$}}}

\def\ed{\end{document}}

\def\Ob{\Omega_b}

\def\Ol{\Omega_\Lambda}
\def\Om{\Omega_m}

\def\ob{\omega_b}

\def\od{\omega_d}

\def\ns{n_s}

\def\beq#1{\begin{equation}\label{#1}}
\def\eeq{\end{equation}}
\def\beqa#1{\begin{eqnarray}\label{#1}}
\def\eeqa{\end{eqnarray}}
\def\eq#1{equation~(\ref{#1})}
\def\Eq#1{Equation~(\ref{#1})}
\def\eqn#1{~(\ref{#1})}

\def\tr{\hbox{tr}\>}

\def\deltah{\widehat{\delta}}
\def\nbar{{\bar n}}

\def\ith{i^{\rm th}}

\def\f{{\bf f}}
\def\k{{\bf k}}
\def\r{{\bf r}}
\def\rhat{\widehat{\bf r}}
\def\ahat{\widehat{\bf a}}
\def\vb{{\bf b}}
\def\p{{\bf p}}
\def\phat{\widehat\p}
\def\ph{\widehat p}
\def\q{{\bf q}}
\def\r{{\bf r}}
\def\v{{\bf v}}
\def\x{{\bf x}}
\def\y{{\bf y}}

\def\B{{\bf B}}
\def\C{{\bf C}}

\def\F{{\bf F}}
\def\I{{\bf I}}

\def\SS{{\bf\Sigma}}

\def\M{{\bf M}}
\def\N{{\bf N}}
\def\P{{\bf P}}

\def\W{{\bf W}}

\def\S{{\bf S}}

\def\nmonte{N_{\rm monte}}

\def\psih{\widehat{\psi}}

\def\M{{\bf M}}

\def\rh{\widehat{\bf r}}

\def\tr{\hbox{tr}\>}
\def\dV{{d^3k\over (2\upi)^3}}
\def\dag{^\dagger}

\def\ngal{N_{\rm gal}}

\def\nx{{N_x}}

\def\l{\ell}
\def\lmax{\l_{\rm max}}
\def\lcut{\l_{\rm cut}}

\def\dt{\delta}
\def\dl{g}
\def\xh{\widehat{\x}}
\def\Px{P_\times}
\def\Pgg{P_{\rm gg}}
\def\Pgv{P_{\rm gv}}
\def\Pvv{P_{\rm vv}}

\def\im{{\rm i}}	

\def\tr{\hbox{tr}\,}

\def\ith{i^{th}}

\documentclass[useAMS]{mn2e}

\font\japit = cmti10 at 10truept
\input{epsfig.sty}

\title
     [Power Spectrum of 2dFGRS]
{\vglue-3.0truecm
\centerline{\japit Submitted to  MNRAS December 2 2001, revised April 14 2002}
\vglue 2.5truecm
\noindent
The Power Spectrum of Galaxies in the 2dF 100k Redshift Survey
\author[M. Tegmark, A. J. S. Hamilton \& Y. Xu]
     {Max Tegmark$^1$, Andrew J. S. Hamilton$^2$ \& Yongzhong Xu$^1$ \\
	$^1$Dept. of Physics, Univ. of Pennsylvania, Philadelphia, PA 19104, USA;
	max@physics.upenn.edu; http:$/\!/$www.hep.upenn.edu/$\sim$max/ \\
	$^2$JILA and Dept.\ Astrophysical \& Planetary Sciences,
	Box 440, U. Colorado, Boulder CO 80309, USA; \\
	\ Andrew.Hamilton@colorado.edu; http:$/\!/$casa.colorado.edu/$\sim$ajsh/}
}

\newcommand{\simpropto}{\!\!\begin{array}{c} {\propto} \\
                  [-1.7ex] \sim \end{array}\!\!}
\renewcommand{\theenumi}{(\roman{enumi})}
\begin{document}

\maketitle

\begin{abstract}
We compute the real-space power spectrum and the redshift-space distortions 
of galaxies in the 2dF 100k galaxy redshift survey using 
pseudo-Karhunen-Lo\`eve eigenmodes and the stochastic bias
formalism.
Our results agree well with those published by the 2dFGRS team, 
and have the added advantage of 
producing easy-to-interpret uncorrelated minimum-variance measurements of the 
galaxy-galaxy, galaxy-velocity and velocity-velocity power spectra 
in 27
$k$-bands, with
narrow and 
well-behaved window functions in the range
$0.01\hperMpc < k < 0.8\hperMpc$. 
We find no significant detection of baryonic wiggles, although our results
are consistent with a standard flat $\Omega_\Lambda=0.7$ ``concordance'' model 
and previous tantalizing hints of baryonic oscillations.
We measure the galaxy-matter correlation 
coefficient $r > 0.4$ 
and the redshift-distortion parameter  
$\beta=0.49\pm 0.16$ for $r=1$ 
($\beta=0.47\pm 0.16$ without finger-of-god compression). 
Since this is an apparent-magnitude limited sample,
luminosity-dependent bias may cause a slight red-tilt in the
power spectum.
A battery of systematic error tests
indicate that the
survey is not only impressive in size, but also unusually clean,
free of systematic errors at the level to which our tests are sensitive.
Our measurements and window functions are available at 
$http://www.hep.upenn.edu/\sim max/2df.html$ together with 
the survey mask, radial selection function and uniform 
subsample of the survey that we have constructed.
\end{abstract}


\section{Introduction}

Three-dimensional maps of the Universe provided by
galaxy redshift surveys place powerful constraints on 
cosmological models, which has motivated ever more ambitious
observational efforts such as the 
the CfA/UZC (Huchra {\etal} 1990; Falco {\etal} 1999),
LCRS (Shechtman {\etal} 1996) and PSCz (Saunders {\etal} 2000)
surveys, each well in excess of $10^4$ galaxies.
This has been an exciting year in this regard, with 
early results released from two even more ambitions projects;
the AAT two degree field galaxy redshift survey (2dFGRS; 
Colless {\etal} 2001) and the Sloan Digital Sky Survey 
(SDSS; 
York {\etal} 1999), which aim for 
250{,}000 and 1 million galaxies, respectively.

Analysis of the first  
147{,}000 2dFGRS galaxies (Peacock {\etal} 2001; Percival {\etal} 2001; Norberg {\etal} 2001a; Madgwick {\etal} 2001) 
and the first 
29{,}000 SDSS galaxies (Zehavi {\etal} 2002)
have supported a flat dark-energy
dominated cosmology, as have angular clustering analyses of the 
parent catalogs underlying the 2dFGRS (Efstathiou \& Moody 2001)
and SDSS 
(Scranton {\etal} 2002; Connolly {\etal} 2002; Tegmark {\etal} 2002; 
Szalay {\etal} 2002; Dodelson {\etal} 2002).
Tantalizing evidence for baryonic wiggles in the galaxy power spectrum
has been discussed (Percival {\etal} 2001; Miller {\etal} 2001),
and cosmological models have been constrained in conjunction with
cosmic microwave background (CMB) data (Efstathiou {\etal} 2002).

The 2dFGRS team has kindly made their first 102{,}000 redshifts publicly 
available. Given the huge effort involved in creating this state-of-the-art
sample, it is clearly worthwhile to subject it to an independent power spectrum
analysis. This is the purpose of the present paper, focusing on 
large $(k\simlt 0.3\hperMpc)$ scales.
Since the cosmological constraints from galaxy surveys are only 
as accurate as our modeling of bias, extinction, integral 
constraints, geometry-induced power smearing 
and other real-world
effects, we will employ a number of recently developed techniques for
tackling these issues. Compared with the
solid and thorough
analysis by
the 2dFGRS team in Peacock {\etal} (2001) and Percival {\etal} (2001), our
main improvements will be in the following areas:
\begin{itemize}
\item By using an approach based on information theory, involving 
pseudo-Karhunen-Lo\`eve eigenmodes, quadratic estimators and 
Fisher matrix decorrelation, we are able to produce uncorrelated measurements
of the
linear
power spectrum with minimal error bars and 
quite narrow window functions. This allows the power spectrum to be 
plotted in an easy-to-interpret model-independent way
and, because of the narrow windows, 
minimizes aliasing from non-linear scales when fitting to linear models.

\item
Using the stochastic bias formalism, we
measure independently not one power spectrum but three,
encoding clustering anisotropy.
On large scales where redshift distortions are linear (Kaiser 1987),
these three curves are the real-space power spectra of 
the galaxies, their velocity divergence (related to the matter density)
and the cross-correlation between the two. On smaller scales,
the information they encode can be extracted using simulations.
\end{itemize}

\begin{figure*} 
\noindent
\centerline{\epsfxsize=17.0cm\epsffile{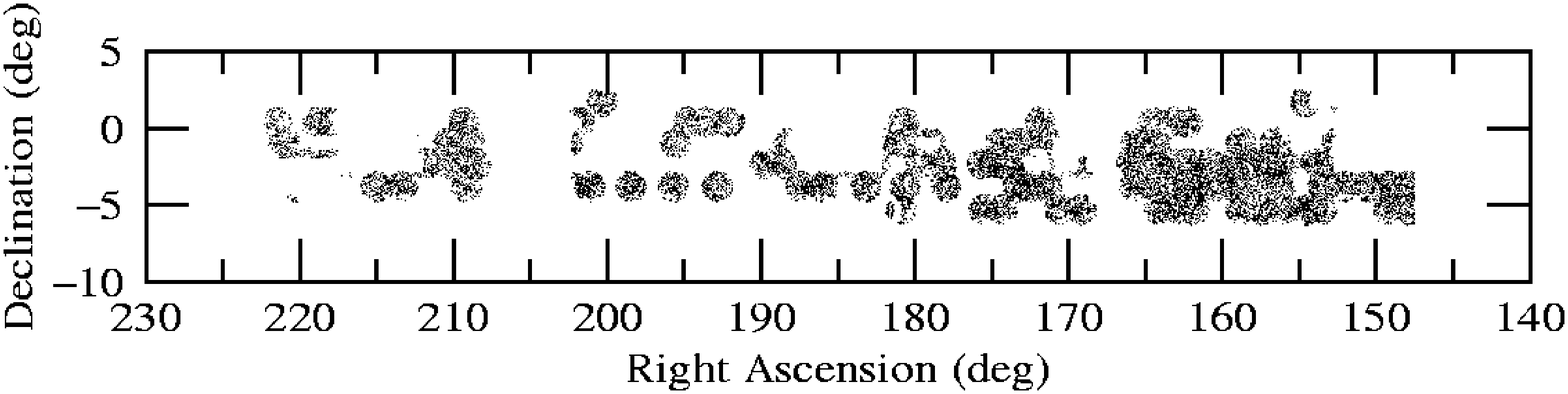}}
\centerline{\epsfxsize=17.0cm\epsffile{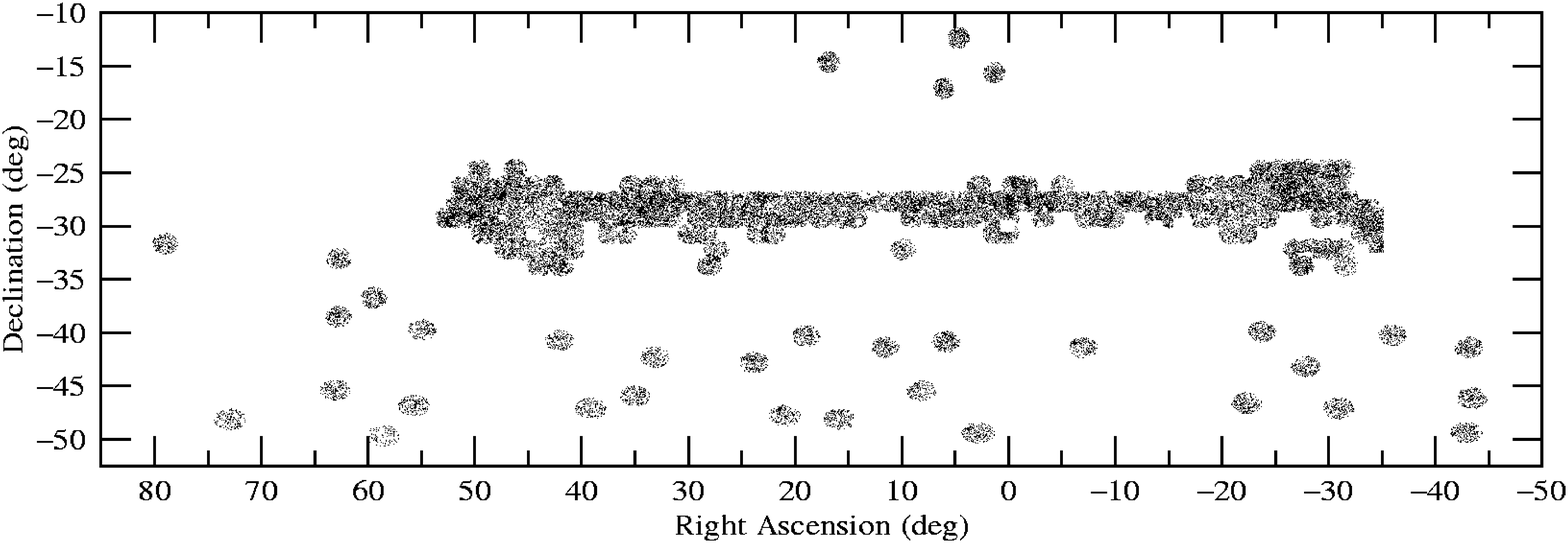}}
\centerline{\epsfxsize=17.0cm\epsffile{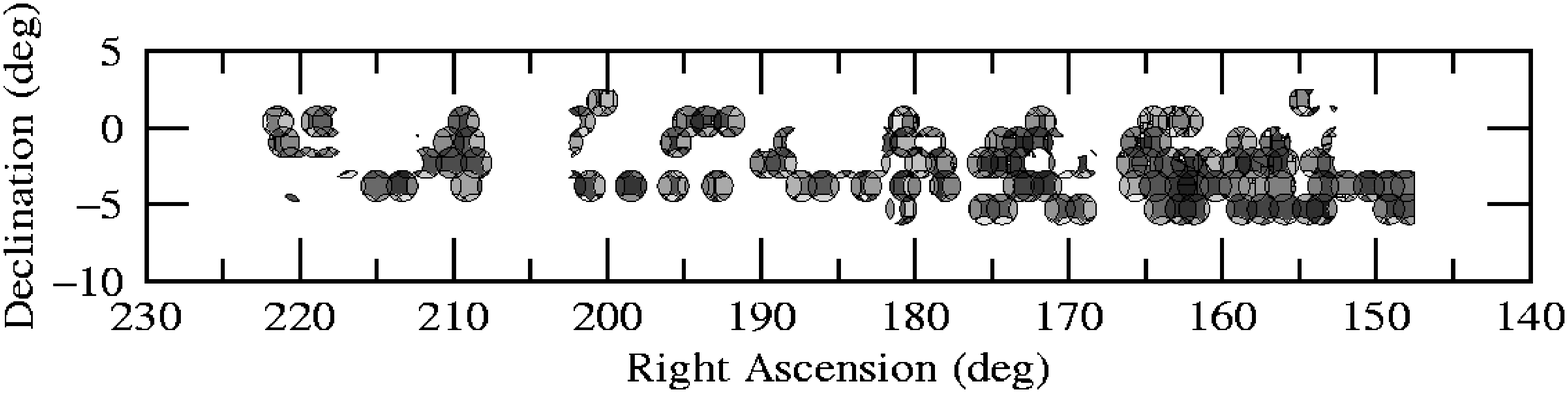}}
\centerline{\epsfxsize=17.0cm\epsffile{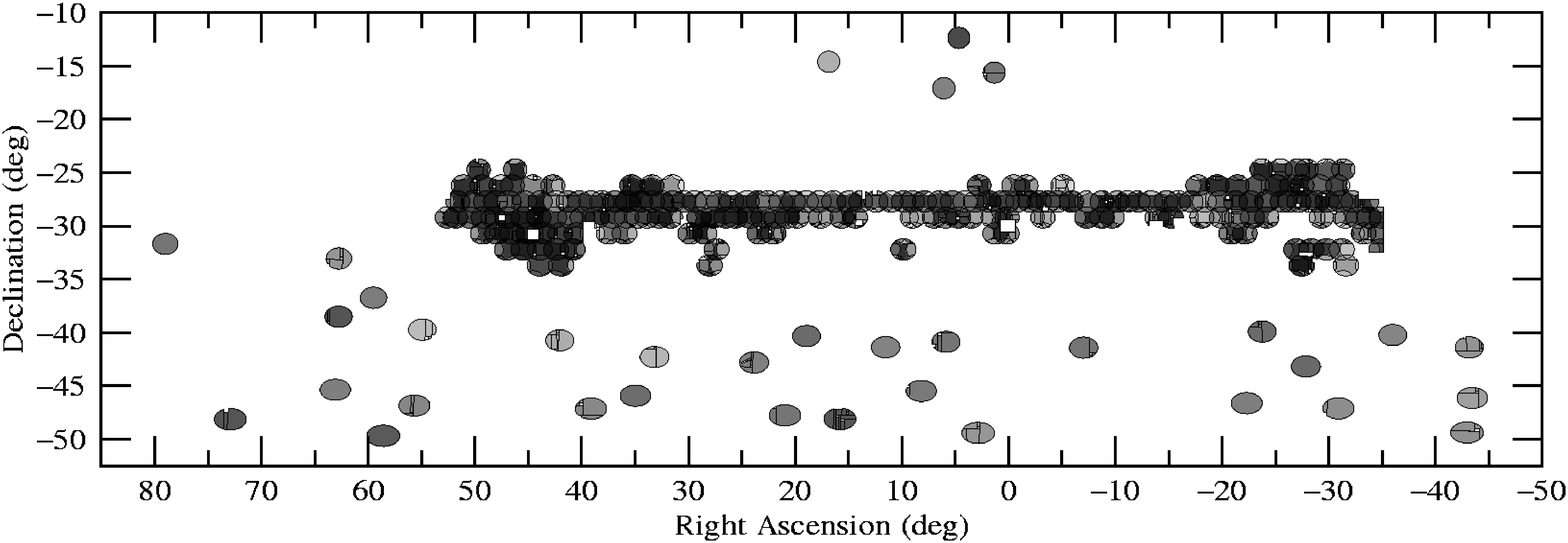}}
\caption[1]{\label{maskFig}\footnotesize%
The upper half shows the 59832 2dF galaxies in our 
baseline sample, in equatorial 1950 coordinates.
The lower half shows the corresponding angular mask, 
the relative probabilities that galaxies in various directions
get included. 
}
\end{figure*}

The rest of this paper is organized as follows.
In \sec{DataSec}, we describe the 2dFGRS data used and construct an
easy-to-interpret subsample that is strictly magnitude limited after 
taking various real-world complications into account.
We perform our basic analysis in \sec{AnalysisSec} and report
the results in \sec{ResultsSec}.
 In \sec{SystematicsSec}, we test for 
a variety of systematic errors in \sec{SystematicsSec} and 
quantify the effect of non-linearity and non-Gaussianity on our 
measurements.
In \sec{ConcSec}, we discuss our results, fit to cosmological models
and compare our results with those in the literature.

\section{Data modeling}
\label{DataSec}

The 2dFGRS is described in detail in Colless {\etal} (2001, hereafter C01).
The publicly released 2dFGRS sample consists of 
102{,}426 unique
objects (excluding duplicates),
of which 93{,}843 have survey quality redshifts (quality factor $\ge 3$).
Of these
5{,}131 objects have heliocentric redshifts $z \le 0.002$
and are therefore probable stars,
while a further 240 galaxies lie outside the defined angular
boundaries of the survey
(usually inside a hole in one of the parent UKST fields,
occasionally marginally outside one of the 381 surveyed $2^\circ$ fields).
This leaves a sample of 88{,}472 galaxies with survey quality redshifts.
To do full justice to the quality of this data set
in a power spectrum analysis, 
it is crucial to model accurately the three-dimensional 
selection function $\nbar(\r)$, which gives the expected (not observed)
number density of galaxies as a function of 3D position.
This is the goal of the present section.

As will be described in \sec{AnalysisSec},
our method for measuring the power spectrum requires,
in its current implementation,
that the selection function be separable into the product of an angular part
and a radial part:
\beq{SeparabilityEq}
\nbar(\r) = \nbar(\rh)\nbar(r),
\eeq
where $\r=r\rh$ and $\rh$ is a unit vector.
The angular part $\nbar(\rh)$ may take any value between 0 and 1,
and gives the completeness as a function of position,
\ie, the fraction of all survey-selected 
galaxies for which survey quality redshifts are actually obtained,
while $\nbar(r)$ gives the radial selection function.
Although it would be possible to generalize the method to a
non-separable selection function (by breaking up the
selection function into a sum of piece-wise separable parts),
we have chosen to stick to the simple case of a separable selection
function, for two reasons.
First, although the selection function of the 2dF 100k release is not separable,
it is nearly so
(the survey was originally designed so that it would be),
and the gain from allowing a non-separable selection function has seemed
insufficient to justify the extra complexity.
Second,
as described in Sections~\ref{angularSec} and \ref{radialSec},
we wish to be able to test for possible systematic effects
arising from a misestimate of extinction,
which would cause a purely angular modulation of
density fluctuations,
or from a misestimate of the radial selection function,
which would cause a purely radial modulation of the density.
Such tests are facilitated if the selection function
is separable.

There are two complications that cause slight departures from 
such separability (C01):
\begin{enumerate}
\item The magnitude limit varies slightly across the sky,
because both the photometric calibration
of the parent UKST fields,
and the extinction correction
at each angular position
was improved after the survey had begun. 
\item
Seeing issues lead to lower completeness for faint galaxies,
and weather variations therefore cause the magnitude-dependent 
completeness fraction to vary
in different $2^\circ$ fields.
\end{enumerate}
Below we will eliminate both of these complications with appropriate
cuts on the data set, obtaining a uniform subsample with a 
separable selection function as in \eq{SeparabilityEq}.

\subsection{The basic angular mask}
\label{MaskSec}

In this subsection, we describe our modeling of the angular
mask $\nbar(\rh)$ for the full sample. In subsequent subsections, we
will shrink and re-weight this mask slightly to eliminate
the above-mentioned complications, obtaining the final result
shown in \fig{maskFig}.

Once the 2dFGRS is complete,
it will contain a total of 1192 circular $2^\circ$ fields,
including 450 fields in
a $75^\circ\times 10^\circ$ strip near the North Galactic Pole,
643 fields in
an $85^\circ\times 15^\circ$ strip near the South Galactic Pole, 
and a further 99 fields distributed randomly around the Southern strip.
The various intersections of these fields with each other
yield 7189 non-overlapping intersection regions, referred to as sectors.
Parts of sectors are excluded if they fall outside 
the boundaries of the 314 rectangular UKST plates of the parent APM survey,
or inside one of the holes excised from the plates
in order to eliminate e.g.\ bright stars and satellite trails.
The data release specifies 2024 holes,
of which
1670
lie within, or overlap, those parts of the UKST plates
designated as part of the 2dF survey.

The 100k release is a subset of the survey,
containing data from 381 circular $2^\circ$ fields,
including 39 random fields.
Eventually, when the survey is done, the observed region will be complete,
but in the interim the released fields are variably incomplete,
with a different completeness fraction $\nbar(\rh)$ in each sector,
as described in C01.
 
As part of the 2dFGRS data release,
Peder Norberg and Shaun Cole provide software
that evaluates $\nbar(\rhat)$ in
each of approximately 2.5 million $3^\prime \times 3^\prime$ pixels,
taking all the various complications into account.
However,
we wish to adopt a different angular mask
that admits a separable selection function,
and we also wish to be able to compute the spherical harmonics
of the angular mask using the fast, analytic method described in
Appendix~A of Hamilton (1993).
We therefore use a more explicit geometric (not pixellized)
specification of the mask, described immediately below.

All field, plate and hole boundaries are simple arcs on the celestial sphere,
corresponding to the intersection of the sphere with some appropriate plane.
This means that any spherical polygon (a field, plate, hole, sector, \etc)
can be defined as the intersection of a set of {\it caps}, where 
a cap is the set of directions $\rh$ satisfying 
$\ahat\cdot\rh > b$ for some unit vector $\ahat$ and some constant 
$b\in[-1{,}1]$.
For instance, a $2^\circ$ field is a single cap, and a rectangular plate is the
intersection of four caps. We define masks such as that the one plotted
in \fig{maskFig} as a list of non-overlapping polygons such that
$\nbar(\rh)$ is constant in each one.
We construct the basic 2dFGRS mask as follows:
\begin{enumerate}

\item We generate a list of
8903 polygons comprised of 7189 sectors
and 1670 holes,
plus 44 polygons defining boundaries of UKST plates.

\item Whenever two polygons intersect, we split them into 
non-intersecting parts, thereby obtaining a longer list of 
12066
non-overlapping polygons. Although slightly tricky in practice,
such an algorithm is easy to visualize: if you draw all boundary lines
on a sphere and give it to your child as a coloring exercise,
using four crayons and not allowing identically colored neighbors,
you would soon be looking at such a list of non-overlapping polygons.

\item We compute the completeness $n(\rh)$ for each of these new polygons,
originally using the Norberg-Cole software,
but subsequently using our own computations, described in the following
subsections.

\item We simplify the result by omitting polygons with zero weight
and merging adjacent polygons that have identical weight.

\end{enumerate}

With the original Norberg and Cole completenesses,
the result is a list of 3765 polygons,
with a total (unweighted) area
of 983 square degrees,
and an effective (weighted) area
$\int\nbar(\rh)d\Omega$ of
of 537 square degrees.

With the revised completenesses described in Section~\ref{angularselfnSec},
there are 3614 polygons,
with an (unweighted) area of
711 square degrees,
and an effective (weighted) area
$\int\nbar(\rh)d\Omega$ of
431 square degrees.
This angular mask,
and the polygons into which it resolves,
are illustrated in Figure~\ref{maskFig}.

Section~\ref{angularselfnSec} explains how we eliminate the two 
above-mentioned complications,
the variations in the magnitude limit,
and the variations in the weather,
so as to produce an angular mask
with the same radial selection function at all points.
The reader uninterested in
such details can safely skip all this, jumping straight 
to \sec{selfuncSec}, 
remembering only the simple bottom line: we create a uniform sample
with 64{,}844 galaxies over 711 square degrees that is complete
down to $b_J$ magnitude 19.27.

\subsubsection{Cutting to a uniform magnitude limit}
\label{maglimSec}

\begin{figure} 
\vskip\smtopskip
\centerline{\epsfxsize=\figsize\epsffile{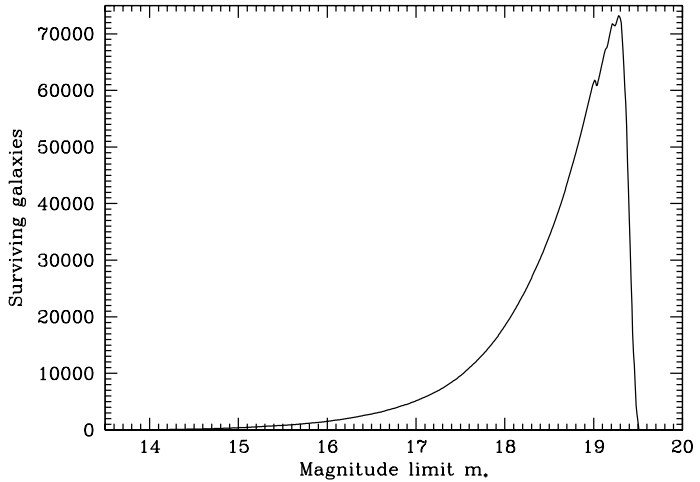}}
\vskip\smbotskip
\caption[1]{\label{survivorsFig}\footnotesize%
Number of galaxies surviving as a function of uniform magnitude cut.
}
\end{figure}

The 2dFGRS aimed to be complete to a limiting $b_J$ magnitude $m=19.45$ after
correction for extinction.
However, the actual limiting magnitude varies slightly across the sky
as described in C01. This is because after the survey began,
there have been improvements in both the
photometric calibrations of the underlying 
parent catalog (Maddox {\etal} 2001, in preparation) 
and in the extinction corrections (Schlegel, Finkbeiner \& Davis 1998).
We eliminate this complication by creating a sub-sample that is complete
down to a slightly brighter limiting magnitude $m_*$, applying the
following two cuts:
\begin{enumerate}
\item Reject all galaxies
whose extinction-corrected magnitude $b_J$ is fainter than $ m_*$.
\item Reject all
sectors whose extinction-corrected magnitude is brighter than $m_*$.
The magnitude limit of a sector is defined in the most
conservative possible fashion:
it is the brightest among all the magnitude limits at the position
of each galaxy and of each Norberg-Cole pixel within the sector.
The extinction at each position is evaluated
using the extinction map of Schlegel, Finkbeiner \& Davis (1988).
\end{enumerate}
\Fig{survivorsFig} shows the number of surviving galaxies as a function 
of $m_*$. As we increase $m_*$, the first cut eliminates 
fewer galaxies whereas the second cut eliminates more galaxies.
The result is seen to be a rather sharply peaked curve, 
taking its maximum for $m_*=19.27$, for which  
66{,}050, or 75~percent, of the 88{,}472 galaxies survive.

The choice $m_*=19.27$ turns out to maximize not only the number 
of galaxies, but the effective survey volume as well.
As the flux cut $F_*$ is made fainter, the depth of the survey 
($\propto F_*^{-1/2}$ in the Euclidean limit) increases, but
the area decreases because there are fewer
sectors complete down to $F_*$.
Therefore the survey volume $\simpropto$(area)$\times F_*^{-3/2}$,
and this also happens to peak for $F_*$ corresponding 
to magnitude 19.27.

\subsection{Angular selection function}
\label{angularselfnSec}

\subsubsection{Modeling the weather}

One of the more time-consuming aspects of our analysis
was modeling another departure from uniformity in the 2dFGRS: 
spatial dependence of the magnitude-dependent incompleteness.
As described by C01,
although the success-rate $P$ for measuring reliable redshifts
(quality $\ge 3$) for targeted
galaxies is in general quite high, it depends on weather.
The poorer the seeing is when a given field is observed,
the lower the success rate. Moreover, this weather modulation 
affects fainter galaxies more than bright ones.
C01
found the success rate to be well fit by an expression of the form
\beq{PmodelEq}
P(F) = \gamma [1 - (\Phi_f/F)^a],
\eeq
where $F$ is the observed flux from the galaxy, 
$\gamma=0.99$, $a=2.5/\ln(10)\approx 1.086$
and $\Phi_f$ is a parameter that is fit for separately for each observed field $f$,
interpretable as the faintest observable flux. 
Note that since this observational selection effect depends only on
magnitude and weather, this issue can be
analyzed and resolved in terms of apparent magnitudes alone,
without explicitly involving redshifts.

The $\Phi_f$-values computed by the 2dFGRS team were not part of
the public release, but it is straightforward to generate
values from the data provided.
Whereas C01 estimated $\Phi_f$ from the observed completeness
fraction for each field, we performed a maximum-likelihood fit
over the fluxes of all objects (galaxies and stars)
targeted for observation in each of the 381 field-nights,
the likelihood being a product of terms $P(F_i)$ for all
successful observations (those yielding a survey-quality redshift),
and terms $[1-P(F_i)]$ for all unsuccessful observations.
Maximizing over 382 parameters
($\Phi_f$ for each of 381 distinct
field-nights $f$,
and a global value of $a$,
with $\gamma$ fixed equal to 1),
we obtain a best fit exponent
$a = 0.96 \pm 0.04$.
Since the exponent is consistent with unity,
we set $a = 1$ for simplicity.
We repeated the analysis with $a$ permitted
to vary separately in each field,
but the likelihood is consistent with constant values.

As a cross-check, we repeated the entire analysis sector-by-sector
instead of field-by-field, obtaining reassuringly similar results.

\subsubsection{Random sampling to a sharp magnitude limit}
\label{muSec}

As mentioned, our power spectrum analysis
requires
a selection function of the separable form of 
\eq{SeparabilityEq}. Yet the discussion above shows
that the shape of the radial selection function 
$\nbar(r)$ varies across the sky,
since the success rate $P_f(F)$ is different for each of the
381
field-nights $f$, as given by \eq{PmodelEq}.

We remedy this problem by sparse-sampling the
galaxies in such a way that the {\it shape} of the success rate
$P(F)$ (as opposed to its amplitude) becomes the same for all
fields. The amplitude variations can then be absorbed into
the angular mask $\nbar(\rh)$, restoring separability.
There are clearly infinitely many ways of doing this --- we wish to
find the way that maximizes the effective volume of the survey
for measuring large scale power.

If we throw away galaxies at random, keeping galaxies in a given 
field $f$ with a probability $p_f$ that depends on their observed flux $F$, 
then the
original success rate $P_f(F)$ for the field from \eq{PmodelEq} gets
replaced by $P_f(F)p_f(F)$.
Our goal then becomes to choose these probabilities $p_f(F)$ such that
\beq{PgoalEq}
P_f(F) p_f(F) = w_f P_*(F),
\eeq
where $P_*(F)$ is the desired uniform, global success rate
and the weights $w_f$ are are scaling factors that will be absorbed into
the angular mask.
Since the functions $P_f(F)$ are known, \eq{PgoalEq} immediately 
specifies how we should choose the probabilities
once the function $P_*$ and the weights have been fixed.
To maximize the number of surviving galaxies, we
want to make $p_f$ and hence $w_f$ as large as possible. 
Since probabilities cannot exceed unity, this implies that the best weights are
\beq{wEq}
w_f = \min_F\left[{P_f(F)\over P_*(F)}\right]
.
\eeq

It remains to choose the target success rate $P_*(F)$.
Since we are interested in large scale power,
our aim is to maximize not so much the number of galaxies,
but rather the effective volume of the survey,
and we must accomplish this goal by adjusting a function $P_*(F)$
of apparent flux $F$.
The way to do this is to choose $P_*(F)$ so as to retain all galaxies
at the faint limit of the survey,
and then to make $P_*(F)$ as large as possible at all other fluxes.
Given that the original $P(F)$ decreases monotonically to fainter fluxes
for all values of the weather parameter $\Phi_f$,
and that $\Phi_f$ includes cases of perfectly observed fields
($\Phi_f = \infty$),
the solution is simply to choose $P_*(F)$ to be constant,
which can be taken to equal 1 without loss of generality,
at all values brighter than the flux limit.

This is delightfully simple and convenient:
it means that the best choice is a pure magnitude-limited
sample with no magnitude incompleteness to keep track of!
The corresponding weights are 
\beq{wEq2}
w_f=\min_F P_f(F) = P_f(F_*) = 1-\Phi_f/F_*
\eeq
where $F_*$ is the flux limit.
The scheme thus keeps all galaxies at the flux limit $F_*$,
and discards a progressively larger fraction of the brighter galaxies
in each sector so as to cancel exactly the magnitude-dependence
of the incompleteness.

The magnitude limit $19.27$ arrived at in the previous
subsection turns out to maximize the number of galaxies not only
before sparse-sampling, but also after sparse-sampling.

The final result is a list of 3614 polygons with associated weights.
available at
http:/$\!$/\discretionary{}{}{}www\discretionary{}{}{}.hep\discretionary{}{}{}.upenn\discretionary{}{}{}.edu/\discretionary{}{}{}$\sim$max/\discretionary{}{}{}2df\discretionary{}{}{}.html
together with the uniform galaxy sample and our power spectrum measurements.
The total area is 711 square degrees, and the effective area
$\int\nbar(\rh)d\Omega$ is 432 square degrees.

\subsection{The radial selection function}
\label{selfuncSec}

After the modeling of angular effects above, it remains
to measure the radial selection function $\nbar(r)$ for the 
uniform sample.
It is important to do this as accurately as possible,
since errors in the selection function translate
into spurious large scale power.

\begin{figure} 
\centerline{\epsfxsize=\figsize\epsffile{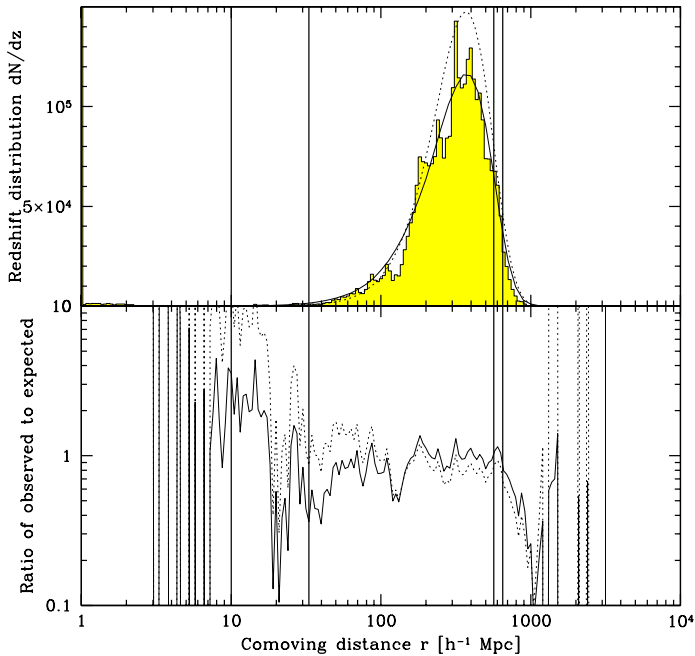}}
\caption[1]{\label{zhistFig}\footnotesize%
The redshift distribution of the galaxies
in our sample is shown both as a histogram (top) and
relative to the expected distribution (bottom), 
in comoving coordinates assuming a flat $\Omega_m=0.3$ cosmology.
The curves correspond to the the radial selection function $\nbar(r)$
employed in our analysis (solid) and 
by C01 (dotted).
The four vertical lines indicate the redshift limits
employed in our analysis ($10\hMpc<r<650\hMpc$)
and where spectral type subsamples are available ($33\hMpc<r<538\hMpc$).
}
\end{figure}

The radial selection function $\nbar(r)$
that results from the analysis described immediately below
is shown in \fig{zhistFig}.

In addition to imposing a faint magnitude limit of $b_J = 19.27$,
we follow the advice of the 2dFGRS team
(Matthew Colless 2001, private communication)
in cutting the survey to a bright limit
of $b_J = 15$.
We use a maximum likelihood method
based on the $C^-$ method of Lynden-Bell (1971),
which assumes that luminosity is uncorrelated with position.
We generate an initial approximation to the selection function
using a continuous version of the Turner (1979) method,
which yields the exact maximum likelihood solution
for the case of a survey with a sharp faint flux limit.
The Turner method has the merit of being exceedingly fast
(less than one CPU second),
but it works only if the survey is flux-limited at one end
(e.g. the faint end).
Starting from the Turner solution,
we use an iterative method designed to converge towards
the exact maximum likelihood solution for the selection function,
which can be shown (Hamilton \& Tegmark 2002) to be a step function
with steps at the limiting distance of each of the
$\sim 60{,}000$ galaxies in the sample.
To implement the Bayesian prejudice that the selection function
should be smooth, we interpolate the resulting 60{,}000-point
function at $\sim 500$ points, through which we pass a cubic spline.

We follow the 2dFGRS team in 
assuming a flat $\Ol=0.7$ cosmology when converting 
redshifts to comoving distances $r$.
We transform the galaxy positions
into the Local Group frame
assuming that the solar motion relative to the Local Group
is 306 km/s toward $l=99^\circ$, $b=-4^\circ$
(Courteau \& van den Bergh 1999).
We model $k$-corrections and luminosity evolution ($\varepsilon$-corrections) 
together as a power law luminosity evolution $\propto (1+z)^\kappa$
with exponent $\kappa = -0.7$.
This exponent
was chosen so as to make the comoving density shown
in the lower panel in \fig{zhistFig} as flat as possible, 
\ie, by assuming minimal evolution in the comoving number density of galaxies.
Similar results have been reported by Cole {\etal} (2001), C01, 
Cross {\etal} (2001) and Madgewick {\etal} (2001) and Norberg {\etal} (2001b).
The slight differences between our $\nbar(r)$ and that of  C01 seen \fig{zhistFig}
are due to our different methods for estimating this function from the data, and 
below we find that they do not have a major impact on the final power spectrum. 

We truncate the sample radially by eliminating objects with
$r<10\hMpc$ (to eliminate stellar contamination) and 
$r>650\hMpc$ (where \fig{zhistFig} shows evidence of incompleteness). 
This leaves 59832 galaxies in the sample.

\section{Method and basic analysis}
\label{AnalysisSec}

In this section, we analyze the uniform galaxy sample described
in the previous section, measuring the power spectrum
and redshift space distortions of the galaxy density field.
We adopt the matrix-based approach described
in Tegmark {\etal} (1998, hereafter THSVS98), using the
mode expansion of Hamilton \& Culhane (1996) and
including the stochastic bias formalism.
Our analysis consists of the following five steps:
\begin{enumerate}
\item Finger-of-god compression
\item Pseudo-Karhunen-Lo\`eve compression
\item True Karhunen-Lo\`eve expansion
\item Quadratic band-power estimation
\item Fisher decorrelation and flavor disentanglement
\end{enumerate}
We will now describe these steps in more detail.
We will see that step (iii) is not required in practice, and we use it only 
for systematics tests.

\subsection{Step 1: Finger-of-god compression}
\label{fogSec}

\begin{figure*} 
\vskip-6cm
\centerline{\hglue0.5cm\epsfxsize=19.5cm\epsffile{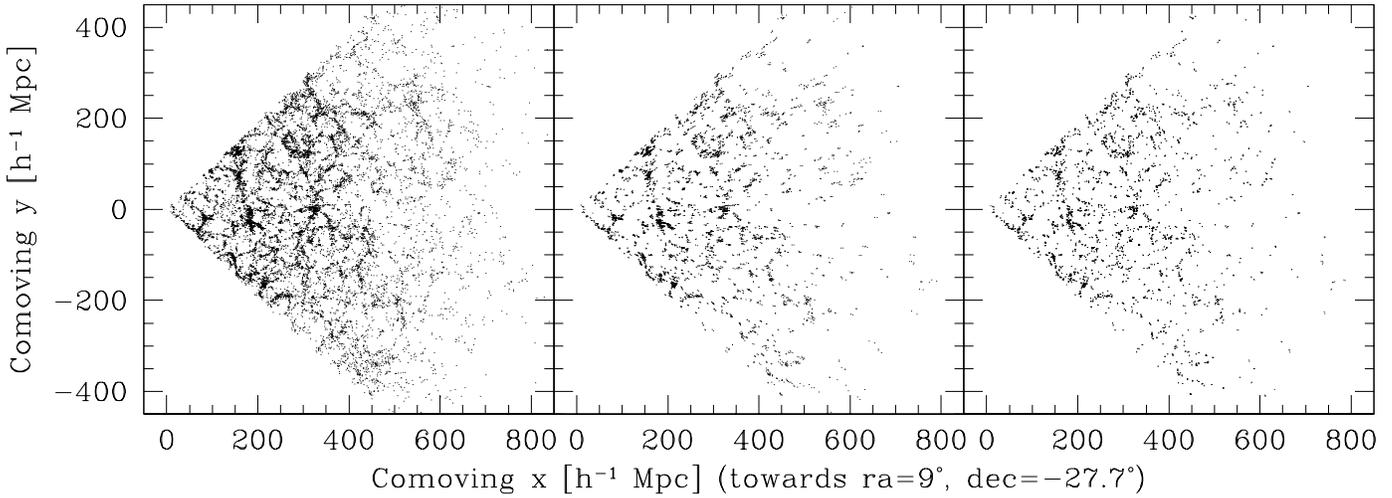}}
\vskip-6cm
\caption[1]{\label{fogFig}\footnotesize%
The effect of our Fingers-of-god (FOG) removal is shown in the
southern slice $\delta=-27.7^\circ$,  $-35^\circ<RA<53^\circ$.
The slice has thickness $2^\circ$ and has been rotated to lie in the
plane of the page. 
From left to right, the panels show all 15{,}055 galaxies in the slice,
the 6{,}211 that are identified as belonging to 
FOGs (with density threshold 100) 
and the same galaxies after FOG compression, respectively.
}
\end{figure*}

Since our analysis uses the linear Kaiser approximation 
for redshift space distortions, it is crucial that we are able to empirically
quantify our sensitivity to the so-called finger-of-god (FOG) effect
whereby radial velocities in virialized clusters make them appear 
elongated along the line of sight. 
We therefore start our analysis by compressing (isotropizing) 
FOGs, as illustrated in \fig{fogFig}.
The FOG compression involves a tunable threshold density,
and in \sec{nonlinearSec} below we will study how the final 
results change as we gradually change this threshold to include 
lesser or greater numbers of FOGs.

We use a standard friends-of-friends algorithm,
in which two galaxies are considered friends, therefore
in the same cluster,
if the density windowed through an ellipse 10 times longer
in the radial than transverse directions,
centered on the pair,
exceeds a certain overdensity threshold.
To avoid linking well-separated galaxies in deep,
sparsely sampled parts of the survey,
we impose the additional constraint that friends
should be closer than
$r_{\perp\max} = 5 \, h^{-1} \Mpc$ in the transverse direction.
The two conditions are combined into the following
single criterion:
two galaxies separated by $r_\parallel$ in the radial
direction and by $r_\perp$ in the transverse direction
are considered friends if
\begin{equation}
\left[(r_\parallel / 10)^2 + r_\perp^2\right]^{1/2}
\le
\left[ (3/4\upi) \upi \nbar ( 1 + \delta_c ) + r_{\perp\max}^{-3} \right]^{-1/3}
\end{equation}
where $\nbar$ is the selection function (geometrically averaged)
at the position of the pair,
and $\delta_c$ is an overdensity threshold.
Note that $\delta_c$ represents not the overdensity of the pair
as seen in redshift space, but rather the overdensity of the pair
after their radial separation has been reduced by a factor of 10.
In other words, $\delta_c$ is intended to approximate the
threshold overdensity of a cluster in real space,
not the overdensity of the elongated FOG seen in redshift space.
Having identified a cluster by friends-of-friends in this fashion,
we measure the dispersion of galaxy positions about the center
of the cluster in both radial and transverse directions.
If the 1-dimensional radial dispersion exceeds the transverse
dispersion, then the cluster is deemed a FOG,
and the FOG is then compressed radially so that the cluster becomes round,
that is, the transverse dispersion equals the radial dispersion.
We perform the entire analysis five times, using 
progressively more aggressive compression with density cutoffs
$1{+}\delta_c=\infty$, 200, 100, 50 and 25, respectively.
The infinite threshold
$1{+}\delta_c=\infty$ corresponds to no compression at all.

\fig{fogFig} illustrates FOG compression with threshold density
$1{+}\delta_c=100$, which is the baseline case adopted in this paper.
It corresponds to fairly aggressive 
FOG removal since the overdensity of a cluster
is around 200 at virialization and rises as the Universe expands
and the background density continues to drop.

\subsection{Step 2: Pseudo-KL pixelization}

\label{PKLsec}

\begin{figure} 
\centerline{\epsfxsize=\figsiz\epsffile{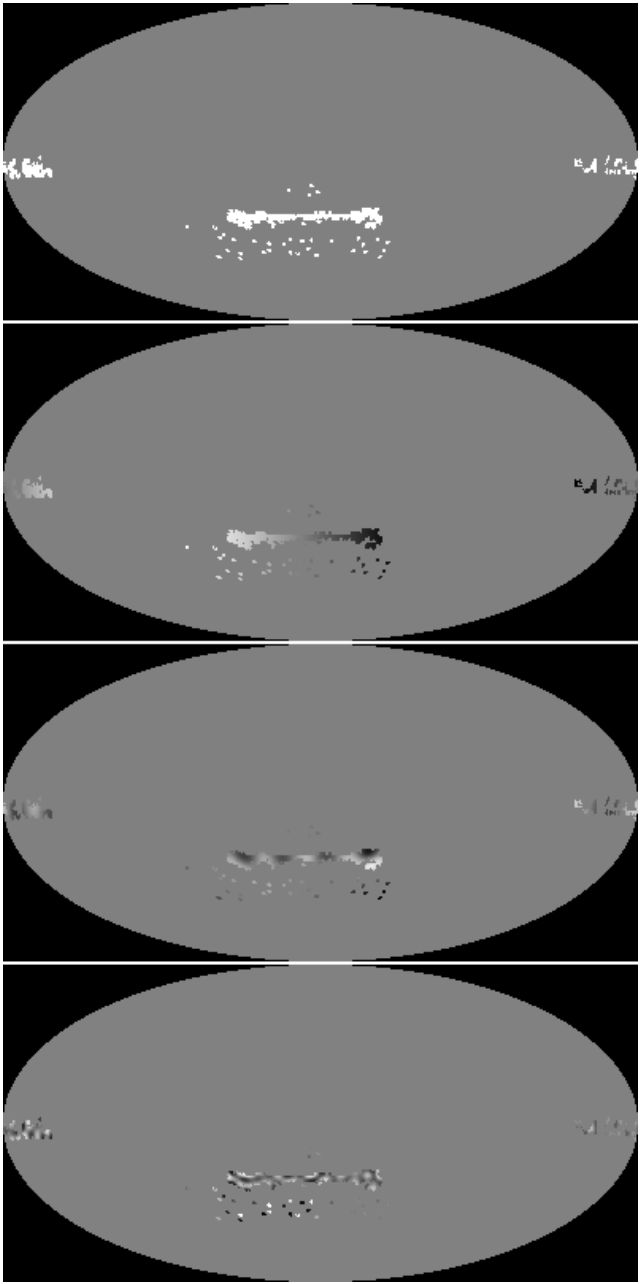}}
\caption[1]{\label{angularmodesFig}\footnotesize%
A sample of four angular pseudo-KL (PKL) modes are shown
in Hammer-Aitoff projection in equatorial coordinates, 
with grey representing zero weight,
and lighter/darker shades indicating positive/negative weight, respectively.
From top to bottom, they are
angular modes 1 (the mean mode), 3, 20 and 106, and
are seen to probe successively smaller angular scales.
}
\end{figure}

\begin{figure} 
\centerline{\epsfxsize=\figsiz\epsffile{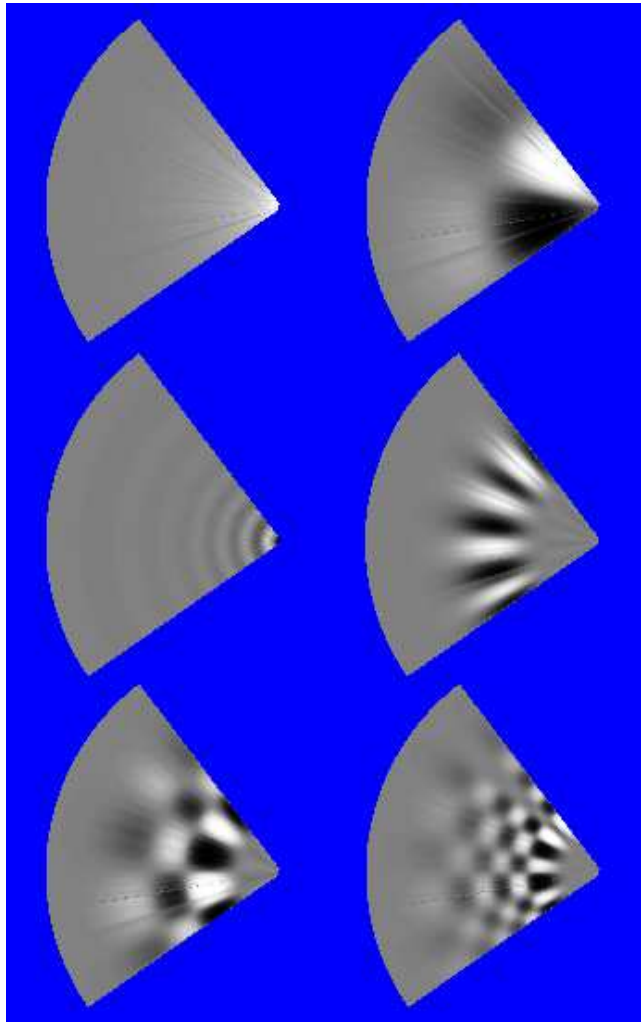}}
\caption[1]{\label{slicemodesFig}\footnotesize%
A sample of six pseudo-KL modes are shown
in the plane of the southern 2dF slice with 
$\delta=-27.7^\circ$,  $-35^\circ<RA<53^\circ$.
Grey represents zero weight,
and lighter/darker shades indicate positive/negative weight, respectively.
From left to right, top to bottom,
these are modes 1 (the mean mode), 14, 104, 148, 58 and 178,
and are seen to probe successively smaller scales.
Those in the middle panel are examples
of purely radial (left) and purely angular (right) modes.
}
\end{figure}

\begin{figure} 
\vskip\smtopskip
\centerline{\epsfxsize=\figsize\epsffile{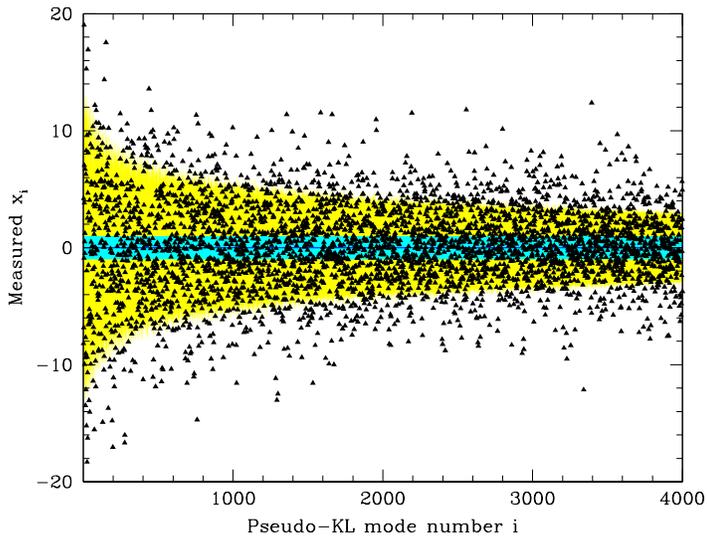}}
\vskip\smbotskip
\caption[1]{\label{xFig}\footnotesize%
The triangles show the 4{,}000 elements $x_i$ of the data vector $\x$
(the pseudo-KL expansion coefficients) 
for the baseline galaxy sample.
If there were no clustering in the survey, merely shot noise, 
they would have unit variance, and about $68\%$ of them would be
expected to lie within the blue/dark grey band.
If our prior power spectrum were correct,
then the standard deviation would be larger, as indicated by the 
shaded yellow/light grey band. 
}
\end{figure}

The raw data consists of $\ngal=59{,}832$ three-dimensional
vectors $\r_\alpha$, $\alpha=1,...,\ngal$, giving the
measured positions of each galaxy in redshift space.
As in THSVS98, we
define the density in $\nx$ 
``pixels'' $x_i$, $i=1,...,\nx$ by 
\beq{xDefEq}
x_i \equiv\int {n(\r)\over\nbar(\r)}\psi_i(\r) d^3r
\eeq
for some set of functions $\psi_i$
and work with the $\nx$-dimensional data vector $\x$ instead of the 
the $3\times\ngal$ numbers $\r_\alpha$.
Although these are perhaps more aptly termed ``modes'' since we will choose
quite non-local functions $\psi_i$, we will keep referring to them as pixels 
to highlight the useful analogy with CMB map analysis. 

Galaxies are
(from a cosmological perspective)
delta-functions in space,
so the integral in equation~(\ref{xDefEq}) reduces to a discrete
sum over galaxies.
We do not rebin the galaxies spatially,
which would artificially degrade the resolution.
It is convenient to isolate the mean density into a single mode
$\psi_1(\r) = \nbar(\r)$,
with amplitude
\beq{meanDefEq1}
x_1=\int n(\r) d^3 r=\ngal,
\eeq
and to arrange all other modes to have zero mean
\beq{meanDefEq2}
\expec{x_i} = \int\psi_i(\r) d^3r = 0
\quad(i \neq 0).
\eeq
The covariance matrix of the vector $\x$ of amplitudes is
a sum of noise and signal terms
\beq{xCovEq}
\expec{\Delta\x\Delta\x^t}=\C\equiv\N+\S,
\eeq
where
the shot noise covariance matrix is given by
\beq{NdefEq}
\N_{ij} = \int {\psi_i(\r)\psi_j(\r)\over\nbar(\r)}d^3 r
\eeq
and the signal covariance matrix is 
\beq{SdefEq}
\S_{ij} = \int\psih_i(\k)\psih_j(\k)^*
P(k)\dV
\eeq
in the absence of redshift-space distortions.
Here hats denote Fourier transforms and
$\nbar$ is the three-dimensional selection function described in 
\sec{DataSec}, {\ie}, $\nbar(\r)dV$
is the expected (not the observed) number of galaxies in 
a volume $dV$ about $\r$.
$P(k)$ is the power spectrum, which for a random field of density fluctuations
$\delta(\r)$ is defined by
$\expec{\deltah(\k)^*\deltah(\k')}=(2\pi)^3\delta_{\rm Dirac}(\k-\k')$.

As our functions $\psi_i(\r)$, we use the 
{\it pseudo-Karhunen-Lo\`eve (PKL) eigenmodes} defined in
Hamilton, Tegmark \& Padmanabhan (2000; hereafter ``HTP00'').
To provide an intuitive feel for the nature of these modes, 
a sample is plotted in \fig{angularmodesFig}
and \fig{slicemodesFig}. We use these modes because they have the
following desirable properties:
\begin{enumerate}

\item They form a complete set of basis functions probing 
successively smaller scales, so that 
a finite number of them (we use the first 4{,}000) allow 
essentially all information about the density field on large scales 
to be distilled into the vector $\x$.

\item They allow
the covariance matrices 
$\N$ and $\S$ to be  fairly rapidly computed.

\item They are the product of an angular and a radial part, \ie,
take the separable form $\psi_i(\r) = \psi_i(\rh)\psi_i(r)$,
which accelerates numerical computations.

\item A range of potential sources of systematic problems are isolated 
into special modes that are orthogonal to all other modes.
This means that we can test for the presence of such problems by looking 
for excess power in these modes, and immunize against their effects
by discarding these modes. 
\end{enumerate}

  We have four types of such special modes:
  \begin{enumerate}
  \item 
   The very first mode is the mean density, 
   $\psi_1(\r)=\nbar(\r)$.
   The mean mode is used in determining the maximum likelihood
   normalization of the selection function,
   but is then discarded from the analysis,
   since it is impossible to measure the fluctuation of the mean mode.
   The idea of solving the so-called integral constraint problem 
   by making all modes orthogonal to the mean
   goes back to Fisher {\etal} (1993). 
   \item 
   Modes 2-5 are associated with the motion of the Local Group
   through the Cosmic Microwave Background
   at 622~km/s towards (B1950 FK4) RA = $162^\circ$, Dec = $-27^\circ$
   (Lineweaver et al.\ 1996;
   Courteau \& van den Bergh 1999).
   In the angular direction, these Local Group modes are monopole and dipole
   modes multiplied by the angular mask, while in the radial direction
   they take the form specified by equation~(4.42) of Hamilton (1997c).
   Mode 2 is a pure monopole mode (multiplied by the angular mask),
   and is present because the survey is not all-sky.
   The other three Local Group modes are dipole modes with admixtures
   of the Local Group monopole mode 2,
   such as to make them orthogonal to the mean mode 1.
   \item 
   Purely radial modes (for example mode 104 in \fig{slicemodesFig})
   are to first order the only ones affected by mis-estimates of
   the radial selection function $\nbar(r)$.
   \item 
   Purely angular modes (for example mode 148 in \fig{slicemodesFig})
   are to first order the only ones affected by misestimates of
   the angular selection function $\nbar(\rh)$, as may result from
   inadequate corrections for, \eg, extinction, the variable magnitude limit,
   the variable magnitude completeness or photometric zero-point offsets.
   \end{enumerate}
   
As described in HTP00, the modes $\psi_i$ are 
computed in the logarithmic spherical wave basis
(Hamilton \& Culhane 1996),
which are orthonormal eigenfunctions
$Z_{\omega \l m}(\r)
= (2 \upi)^{-1/2} \discretionary{}{}{} e^{- (3/2 + \im \omega) \ln r} \discretionary{}{}{} Y_{\l m} (\hat\r)$
of the complete set of commuting Hermitian operators
\begin{equation}
\label{op}
    \im \left( {\partial \over \partial \ln r} + {3 \over 2} \right) =
  - \im \left( {\partial \over \partial \ln k} + {3 \over 2} \right)
  \ , \quad
  L^2
  \ , \quad
  L_z
  \ .
\end{equation}
Slightly better numerical behavior is obtained by expanding
not $\psi_i(\r)$ itself but rather $\psi_i(\r) / \nbar(r)^{1/2}$
(the denominator is the square root of the radial
part of the selection function only, not the angular part)
in logarithmic spherical waves,
since this mitigates some difficulties that arise from the
fact that the radial selection function $\nbar(r)$ varies
by orders of magnitude.
The merits of working in a basis of spherical harmonics were
first emphasized by
Fisher, Scharf \& Lahav (1994)
and by Heavens \& Taylor (1995).
The advantages of working with logarithmic radial waves
$e^{- (3/2 + \im \omega) \ln r}$,
compared for example to spherical Bessel functions,
are both numerical and physical:
\begin{enumerate}
\item
Numerically, the logarithmic radial wave basis permits rapid transformation
between real, $\omega$, and Four\-ier space using Fast Fourier Transforms.
The transformation is mathematically equivalent to the
Fast Fourier-Hankel-Bessel Transform
FFTL{\sc og} described in Appendix~B of Hamilton (2000).
\item
Physically,
logarithmic radial waves are well matched to real galaxy surveys like the
2dFGRS,
which are finely sampled nearby, and coarsely sampled far away.
\item
The linear redshift distortion operator is diagonal in this basis
(Hamilton \& Culhane 1996).
\end{enumerate}

The logarithmic radial wave basis discretizes naturally
on to a 
logarithmically equispaced grid
(in both real and Fourier space),
and is periodic over a logarithmic interval.
To avoid potential problems of aliasing between small and large scales,
we embed the survey inside a suitably large logarithmic interval of depths,
extending in real space from
$10^{-2} \, h^{-1} \Mpc$ to $10^{4} \, h^{-1} \Mpc$.
As remarked in Section~\ref{selfuncSec},
we truncate the survey to radial depths
10--$650 \, h^{-1} \Mpc$ within this interval.

The dimensionless log-frequency $\omega$
in the radial eigenmode $e^{- (3/2 + \im \omega) \ln r}$
is a radial analogue (in a precise mathematical sense)
of the dimensionless angular harmonic number $\l$.
Similar resolution in the radial and angular directions
is secured by choosing the maximum log-frequency
to be about equal to the the maximum harmonic number,
$\omega_{\rm max} \approx \lmax$.
The maximum log-frequency is related to the radial resolution $\Delta\ln r$ by
$\omega_{\rm max} = \upi / \Delta\ln r$.
We adopt a maximum harmonic number of $\lmax = 40$,
and a radial resolution of 32 points per decade,
so $\Delta \ln r = (\ln 10) / 32$,
giving $\omega_{\rm max} = 43.7$
(the same as in HTP00).
These choices ensure comparable effective resolutions in radial and angular directions.

A maximum angular harmonic number of $\lmax = 40$
gives $(\lmax+1)^2=1681$ spherical harmonics,
while 32 points per radial decade over 6 decades
gives 192 radial modes.
Thus there is a potential pool of
$41^2 \times 192 \approx 320{,}000$ modes
from which we would like to construct Karhunen-Lo\`eve (KL) modes.
The usual way to construct such modes
would be to diagonalize a $320{,}000 \times 320{,}000$ matrix,
but this is evidently utterly intractable numerically.

How do we build the PKL modes in practice?
To make the problem tractable,
we instead proceed hierarchically,
first constructing angular PKL modes,
and then constructing a set of radial PKL modes associated with each angular KL mode.
The procedure is possible because we have required the
selection function to be separable into angular and radial parts,
\eq{SeparabilityEq}.
We refer to the resulting modes as pseudo Karhunen-Lo\`eve (PKL) modes.
The PKL basis contains almost as much information as a true KL basis,
but it circumvents the need to diagonalize an impossibly huge matrix.
Our procedure is the same as that of HTP00.
A different, but similar in spirit, hierarchical approach
to the KL problem has been proposed by Taylor et al.\ (2001).

As we proceed from angular PKL mode to angular PKL mode, 
extending each into
3D PKL modes by computing associated
radial functions, we retain only the $N_x=4000$ PKL modes
with the highest expected signal-to-noise. 
As detailed below, we make this truncation both to 
render the various $N_x\times N_\x$ matrices numerically tractable and
to limit sensitivity to small, nonlinear scales. 
As the signal-to-noise of the angular PKL mode decreases,
fewer and fewer of the associated radial PKL modes make the cut
into the pool of PKL modes.
We stop when 10 successive angular PKL modes have contributed
no new PKL mode.
In practice only 140 of the angular PKL modes actually contribute
to the PKL modes.
The reduction from 1681 to 140 angular modes with little information loss
is possible because the spherical harmonics are 
overcomplete and redundant on the modest fraction of the sky actually
covered by the 2dFGRS.

The orthogonality of the PKL modes to the
mean and the properties of the "special" modes 
are enforced in the construction of the modes. 
We perform the PKL decomposition {\it after} selecting out the special
modes (rather than doing the KL decomposition and then making them
orthogonal to the special modes), since we find that this makes better
PKL modes. 
We do this as decribed in Appendix B of THSVS98, with the complication that
we make the non-special modes {\it exactly} orthogonal to the masked mean
and the masked LG modes, not merely orthogonal up to the finite 
order of the discrete matrices.

The pixelized data vector $\x$ is shown in 
\fig{xFig}. This data compression step has thus distilled 
the large-scale information about the galaxy density field
from $\ngal=59{,}832$ galaxy position vectors into 
4{,}000 PKL-coefficients.
The functions $\psi_i$ are normalized so that 
$\N_{ii}=1$, \ie, so that the shot noise contribution to their 
variance is unity. If there were no cosmological density fluctuations
in the survey, merely Poisson fluctuations, 
the PKL-coefficients $x_i$ would thus have 
unit variance, and about $68\%$ of them would be
expected to lie within the blue/dark grey band.
\Fig{xFig} shows that the fluctuations are considerably larger
than Poisson, especially for the largest-scale modes (to the left),
demonstrating that cosmological density fluctuations are
present, as expected.

\subsection{Step 3: Expansion into true KL modes}

Karhunen-Lo\`eve (KL) expansion (Karhunen 1947) 
was first introduced into large-scale structure analysis
by Vogeley \& Szalay (1996). It has since been applied to the 
Las Campanas redshift survey
(Matsubara {\etal} 1999), the UZC survey (PTH01) and the
SDSS (Szalay {\etal} 2002; Tegmark {\etal} 2002) and has been 
successfully applied to Cosmic Microwave Background data as well,
first by Bond (1995) and Bunn (1995).

\begin{figure} 
\vskip\smtopskip
\centerline{\epsfxsize=\figsize\epsffile{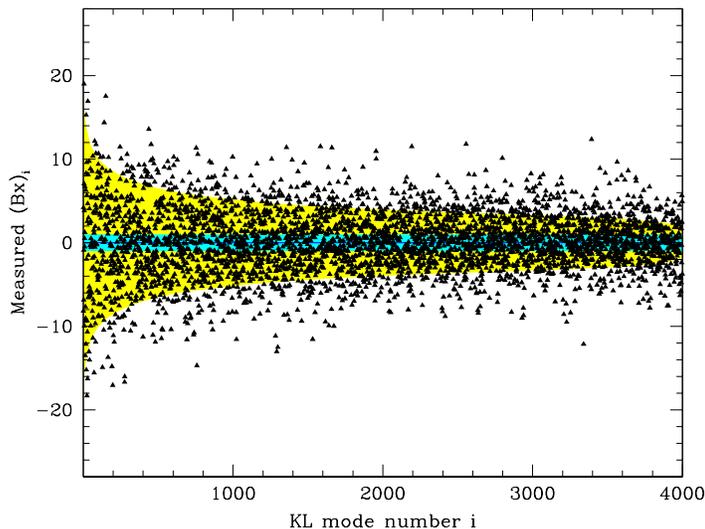}}
\vskip\smbotskip
\caption[1]{\label{WxFig}\footnotesize%
The triangles show the 3999 uncorrelated elements $y_i$ of the transformed data vector 
$\y=\B\x$ (the true KL expansion coefficients) 
for the baseline galaxy sample.
If there were no clustering in the survey, merely shot noise, 
they would have unit variance, and about $68\%$ of them would be
expected to lie within the blue/dark grey band.
If our prior power spectrum were correct,
then the standard deviation would be larger, as indicated by the 
shaded yellow/light grey band. The green/grey curve is the rms of the data points
$x_i$, averaged in bands of width 25, and is seen to 
agree better with the yellow/light grey band than the 
blue/dark grey band.
}
\end{figure}

Given $\x$, $\N$ and $\S$ from the previous section, it is straightforward
to compute the true Karhunen-Lo\`eve (KL) coefficients.
They are defined by
\beq{zDefEq}
\y\equiv\B^t\x,
\eeq
where $\vb$, the columns of the matrix $\B$, are the $\nx$
eigenvectors of the generalized eigenvalue problem
\beq{SNeigenEq}
\S\vb = \lambda \N\vb,
\eeq
sorted from highest to lowest eigenvalue $\lambda$
and normalized so that $\vb\dag\N\vb=\I$.
This implies that
\beq{SNexpecEq}
\expec{y_i y_j}=\delta_{ij}(1+\lambda_i),
\eeq
which means that the transformed data values $\y$ have the desirable
property of being uncorrelated.
In the approximation that the distribution function
of $\x$ is a multivariate Gaussian, this 
also implies that they are statistically independent ---
then $\y$ is merely a vector of independent
Gaussian random variables. 
Moreover, \eq{SNeigenEq} shows that the eigenvalues $\lambda_i$ 
can be interpreted as a signal-to-noise ratio $S/N$.
Since the matrix $\B$ is invertible, the final data set $\y$ clearly retains
all the information that was present in $\x$.
In summary, the KL transformation partitions the information content of 
the original data set $\x$ into $\nx$ chunks that are
mutually exclusive (independent),
collectively exhaustive (jointly retaining all the information), and
sorted from best to worst in terms of their information content.
In most applications, the chief use of KL-coefficients is for data compression,
discarding modes containing almost no information and thereby accelerating 
subsequent calculations.
The KL-coefficients for our dataset are plotted in \fig{WxFig}, and it is 
seen that even the worst coefficients still have non-negligible signal-to-noise,
bearing numerical testimony to the quality of the PKL-modes we have used.
This means that KL-compression would not accelerate our particular analysis,
and we will indeed work directly with the uncompressed data $\x$ in the following
subsections.

Rather, the reason we have computed KL-coefficients
is as an additional check against systematic errors and 
incorrect assumptions, to verify that we modeled not only
the diagonal terms in $\C$ correctly (as seen in \fig{xFig}),
but the off-diagonal correlations as well.
As discussed in many of the above-mentioned KL-papers, 
inspection of the KL-coefficients as in \fig{WxFig} provides yet another 
opportunity to detect suspicious outliers and to check whether the
variance predicted by the prior power spectrum is consistent with the data.
We will provide a detailed test based on the KL-coefficients in \sec{MonteSec}.

\subsection{What we wish to measure: three power spectra, not one}
\label{biasSec}

Before analyzing the $\x$-vector in the following subsections,
let us first discuss precisely what we want to measure.
Cosmological constraints based on galaxy power spectrum
measurements are only as accurate as our
understanding of biasing. We will therefore perform our analysis
in a way that avoids making any assumptions about the relation
between galaxies and matter, as described in 
Tegmark (1998) and HTP00.

Unfortunately, bias is complicated. 
The commonly used assumption that the matter density fluctuations 
$\delta(\r)$ and the galaxy number density fluctuations $g(\r)$
obey
\beq{SimpleBiasEq}
g(\r) = b\,\delta(\r)
\eeq
for some constant $b$ (the bias factor) appears to be violated
in a number of ways.
It has been long known (Davis \& Geller 1976; Dressler 1980) that
$b$ must depend on galaxy type. However, there is also evidence that 
it depends on scale (see {\eg} Mann {\etal} 1997; Blanton {\etal} 1999; Hamilton \& Tegmark 2002 and references therein) and on time
(Fry 1996; Tegmark \& Peebles 1998; Giavalisco {\etal} 1998). 
Finally, there are good reasons to believe that there is 
{\it no} deterministic relation that can replace \eq{SimpleBiasEq}, 
but that bias is inherently somewhat stochastic (Dekel \& Lahav 1999)
--- this has been demonstrated in both simulations
(Blanton {\etal} 2000) and real data (Tegmark \& Bromley 1999).
The term stochastic does of course not imply any randomness in
the galaxy formation process, merely that additional factors
besides density may be important (gas temperature, say).

The good news for our present analysis is that, 
restricting attention to second moments, 
all the information about stochasticity is contained in a single new
function $r(k)$ (Pen 1998; Tegmark \& Peebles 1998). 
Grouping the fluctuations into a two-dimensional vector
\beq{stochbiasxDefEq}
\x\equiv\left({\dt\atop\dl}\right)
\eeq
and assuming nothing except translational invariance,
its Fourier transform $\xh(\k)\equiv\int e^{-i\k\cdot\r}\x(\r)d^3r$ obeys 
\beq{MatrixPowerEq}
\expec{\xh(\k)\xh(\k')^\dagger}=(2\upi)^3\delta^D(\k-\k')
\left(\begin{tabular}{cc}
$P(\k )$ & $\Px(\k )$ \\
$\Px (\k )$ & $\Pgg(\k )$
\end{tabular}\right)
\eeq
for some $2\times 2$ power spectrum matrix that we will denote $\P(\k)$.
Here $P$ is the conventional power spectrum of the mass distribution, 
$\Pgg$ is the power spectrum of the galaxies, and $\Px$ is the cross spectrum. 
It is convenient to rewrite this covariance matrix as 
\beq{PdefEq}
\P(\k) =
P(\k)\left(\begin{tabular}{cc}
$1$ & $b(\k)r(\k)$ \\ $b(\k)r(\k)$ & $b(\k)^2$ \end{tabular}\right)
\eeq
where $b\equiv(\Pgg/P)^{1/2}$ is the bias factor
(the ratio of galaxy and total fluctuations)
and the new function $r\equiv\Px/(P \Pgg)^{1/2}$ is the dimensionless 
correlation coefficient between galaxies and matter.
Note that both $b$ and $r$ generally depend on scale $k$.
The Schwarz inequality shows that the special case 
$r=1$ implies the simple deterministic \eq{SimpleBiasEq},
and the converse is of course true as well.

On large scales where linear perturbation theory is valid,
redshift distortions (Kaiser 1987; Hamilton 1998) conveniently
allow all three of these functions to be measured.
Specifically, the correlation between the observed densities at any two points
depends linearly on these three power spectra: 
\begin{equation}
\label{Pks}
  \begin{array}{r@{\ }c@{\ }c@{\ }c@{\ }l}
    \mbox{galaxy-galaxy power}    &:&\Pgg(k)&=& b(k)^2 P(k) \\
    \mbox{galaxy-velocity power}  &:&\Pgv(k)&=& r(k) b(k) f P(k)\\ 
    \mbox{velocity-velocity power}&:&\Pvv(k)&=& f^2 P(k)\\ 
  \end{array}
\end{equation}
Here $f\approx\Omega_{\rm m}^{0.6}$ is the
dimensionless growth rate for linear density perturbations (see Hamilton 2001).
More correctly, the `velocity' here refers to minus the velocity divergence,
which in linear theory is related to the mass (not galaxy)
overdensity $\delta$ by
$f\delta + \nabla\cdot\v = 0$,
where $\nabla$ denotes the comoving gradient in velocity units.
Note that $\Pgv(k) = f\Px(k)$ and that the parameter $f$ is 
conveniently eliminated by defining $\beta(k)\equiv f/b(k)$, which gives
\beqa{fzapEq}
\Pgv(k)&=&\beta(k)r(k)\Pgg(k),\nonumber\\
\Pvv(k)&=&\beta(k)^2\Pgg(k).
\eeqa

\subsection{Step 4: Quadratic compression into band powers}
\label{qSec}

In this step, we perform a much more radical data compression
by taking certain quadratic combinations of the data vector
that can easily be converted into power spectrum measurements.

\begin{figure} 
\vskip-0.5cm
\epsfxsize=17cm\hglue-2mm\epsffile{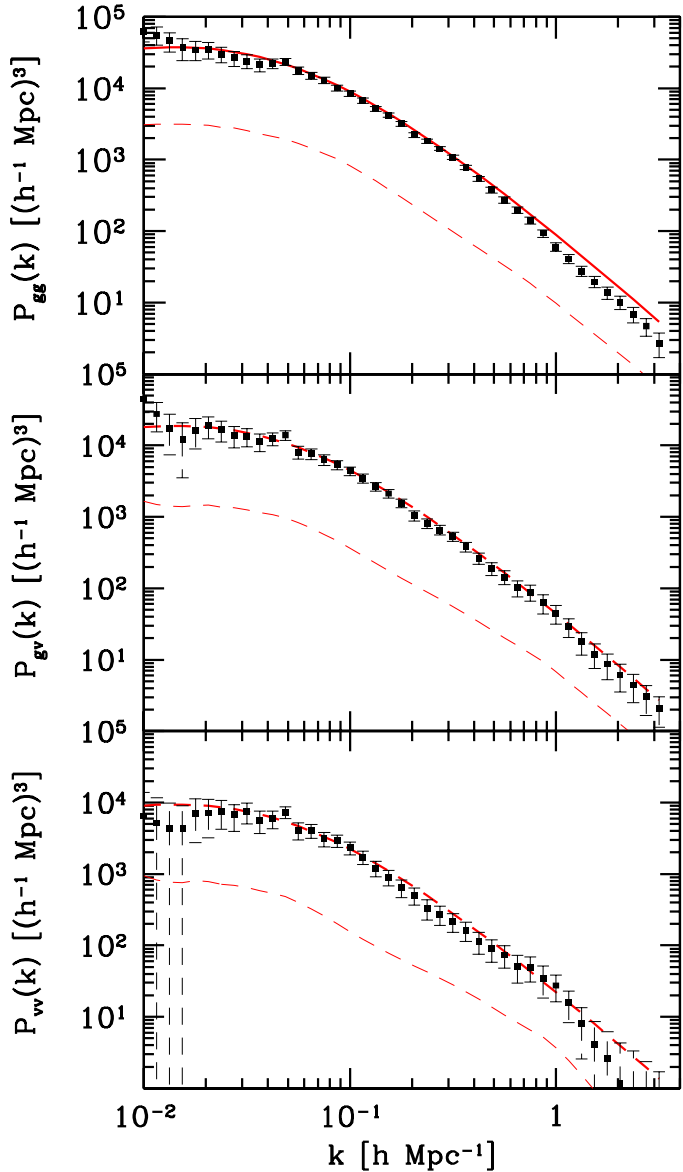}
\vskip-0.5cm
\caption[1]{\label{qFig}\footnotesize%
The 147 quadratic estimators $q_i$, normalized so that their
window functions equal unity and with the shot noise contribution
$f_i$ (dashed curve) subtracted out.
They {\bf cannot} be directly interpreted as power spectrum measurements,
since each point probes a linear combination of 
all three power spectra over a broad range of scales,
typically centered at a $k$-value different than the nominal 
$k$ where it is plotted. Moreover, nearby points are 
strongly correlated, causing this plot to 
overrepresent the amount of information present in the data.
The solid curves show the prior power spectrum used to compute the error bars.
}
\end{figure}

\begin{figure} 
\vskip-1cm
\centerline{\epsfxsize=\figsize\epsffile{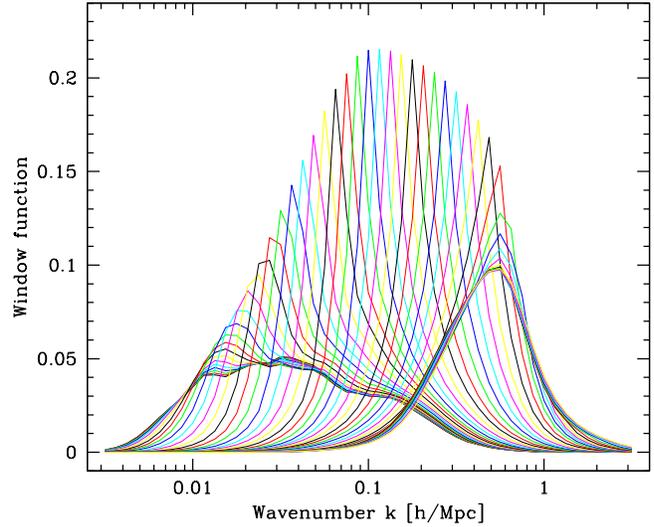}}
\vskip\smbotskip
\caption[1]{\label{Ffig}\footnotesize%
The rows of the $gg$-portion of the Fisher matrix $\F$.
The $\ith$ row typically peaks at the
$\ith$ band, the scale $k$ that the band power 
estimator $q_i$ was designed to probe.
All curves have been renormalized to unit area, 
so the highest peaks indicate the scales the the window 
functions obtained are narrowest.
The turnover in the envelope at $k\sim 0.1\hperMpc$
reflects our running out of information due to 
omission of modes probing smaller scales.
For comparison with the next figure, these are the rows of 
$\W$ when $\M$ is diagonal.  
}
\end{figure}

We parametrize the three power spectra $\Pgg(k)$, $\Pgv(k)$ and $\Pvv(k)$
as piecewise constant functions, each with
49 ``steps'' of height $p_i$, which we term the {\it band powers}.
To avoid unnecessarily jagged spectra, we take
$k^{1.5}P$ rather than $P$ to be constant within each band.
We group these $3\times 49$ numbers into a
147-dimensional vector $\p$.
We choose our 49 $k$-bands to be 
centered on logarithmically equispaced $k$-values
$k_i = 10^{i-41\over 16}\hperMpc$, $i=1,...,49$,
\ie, ranging from $0.00316\hperMpc$ to $3.16\hperMpc$.   
For instance, 
$\Pgg(k) = (k/k_i)^{-1.5}p_i$ for $|\lg k- \lg k_i|<1/32$.
This should provide fine enough 
$k$-resolution to  
resolve any baryonic wiggles and other spectral features
that may be present in the power spectrum.
For instance, baryon wiggles have a characteristic scale of order
$\Delta k\sim 0.1$, so we oversample the first one around $k\sim 0.1$ by
a factor $\Delta k/(k_{26}-k_{25}) \sim 16/\ln 10 \sim 7$.

This parametrization means that we can write
the pixel covariance matrix of \eq{xCovEq} as
\beq{CsumEq}
\C = \sum_{i=0}^{147} p_i \C,{_i},
\eeq
where the derivative matrix 
$\C,{_i}\equiv\partial\C/\partial p_i$
is the contribution from the $\ith$ band.
For notational convenience, we have included the noise term
in \eq{CsumEq} by defining $\C,_{0}\equiv\N$, corresponding
to an extra dummy parameter $p_{0}=1$ giving the shot noise normalization.
As in Hamilton \& Tegmark (2000) and HTP00,
we in practice redefine the parameters $p_i$ to be ratio of the 
actual band power to the prior band power. As long as the prior agrees 
fairly well with the measured result, this has the advantage of giving 
better behaved window functions, as described in Hamilton \& Tegmark (2000).

Our quadratic band power estimates are defined by
\beq{yDefEq2}
q_i\equiv {1\over 2}\x^t\C^{-1}\C_{,i}\C^{-1}\x,
\eeq
$i=0,...,147$.
These numbers are shown in \fig{qFig}, and we group them
together in a $148$-dimensional vector $\q$.
Note that whereas $\x$ (and therefore $\C$) 
is dimensionless, $\p$ has units of power, \ie, volume.
\Eq{yDefEq2} therefore shows that
$\q$ has units of inverse power, \ie, inverse volume.
It is not immediately obvious that the vector $\q$ is a useful 
quantity. It is certainly not the final result (the power
spectrum estimates) that we want, since it does not even 
have the right units. Rather, it is a useful intermediate step.
In the approximation that the pixelized data has a 
Gaussian probability distribution (a good approximation in 
our case because of the central limit theorem, since $\ngal$ is large)
$\q$ has been shown
to retain all the power spectrum information
from the original data set (Tegmark 1997, hereafter ``T97'').
The numbers $q_i$ have the additional 
advantage (as compared with, \eg, maximum-likelihood estimators)
that their properties are easy to compute: their mean 
and covariance
are given by
\beqa{qExpecEq}
\expec{\q}&=&\F\p,\\
\label{qCovarEq}
\expec{\q\q^t}-\expec{\q}\expec{\q}^t &=&\F,
\eeqa
where $\F$ is the {\it Fisher information matrix} (Tegmark {\etal} 1997)
\beq{GaussFisherEq}
\F_{ij} = {1\over 2}\tr\left[\C^{-1}\C_{,i}\C^{-1}\C_{,j}\right].
\eeq
Quadratic estimators were first derived for galaxy survey
applications (Hamilton 1997ab). They were accelerated and 
first applied to CMB analysis (T97; Bond, Jaffe \& Knox 2000). 

In conclusion, this step takes the vector $\x$ and its covariance matrix
$\C$ from \fig{xFig} and compresses it 
into the smaller vector $\q$ and its covariance matrix
$\F$, illustrated in \fig{qFig} and \fig{Ffig}.
Although \eq{qExpecEq} shows that we can obtain unbiased estimates of the 
true powers $\p$ by computing $\F^{-1}\q$, there are better options,
as will be described in the next subsection.

\subsection{Step 5: Fisher decorrelation and flavor disentanglement}

Let us first eliminate the shot-noise dummy parameter $p_0$, 
since we know its value.
We define $\f$ to be the $0^{th}$ column 
of the Fisher matrix defined above
($f_i\equiv \F_{i0}$)
and restrict the indices $i$ and $j$ to run from
$1$ to $147$ from now on, so 
$\f$, $\q$ and $\p$ are $147$-dimensional vectors
and $\F$ is a $147\times 147$ matrix.
Since $p_0=1$, \eq{qExpecEq} then becomes
$\expec{\q}=\F\p + \f$.

We now define a vector of shot noise corrected band power estimates
\beq{phatDefEq}
\phat\equiv\M(\q-\f),
\eeq
where 
$\M$ is some matrix whose rows are 
normalized so that the rows of
$\M\F$ sum to unity. 
Using equations\eqn{qExpecEq} and\eqn{qCovarEq}, this gives
the mean and covariance
\beqa{pExpecEq}
\expec{\phat}&=&\W\p,\\
\label{pCovarEq}
\SS &\equiv& \expec{\phat\phat^t}-\expec{\phat}\expec{\phat}^t =\M\F\M^t,
\eeqa
where $\W\equiv\M\F$. We will refer to the rows of $\W$ as 
window functions,
since they sum to unity and \eq{pExpecEq} shows that
$\ph_i$ probes a weighted average of the true band powers $p_j$,
the $\ith$ row of $\W$ giving the weights.

\subsubsection{Correlated, anticorrelated and uncorrelated band powers}

For the purpose of fitting models $\p$ to our measurements $\phat$, 
we are already done --- the last two equations tell us how to compute $\chi^2$
for any given $\p$, and the result 
\beq{chi2Eq}
\chi2 = (\phat-\expec{\phat})^t\SS^{-1}(\phat-\expec{\phat})^t
\eeq
is independent of the choice of $\M$.
However, since one of the key goals of our analysis is to provide
model-independent measurement of the three power spectra, 
the choice of $\M$ is crucial. Ideally, we would like
both uncorrelated error bars (diagonal $\SS$) and 
well-behaved
(narrow, unimodal and non-negative) window functions $\W$ that
do not mix the three power spectra.

There are a number of interesting choices of $\M$ that each have their
pros and cons (Tegmark \& Hamilton 1998; Hamilton \& Tegmark 2000).
The simple choice where $\M$ is diagonal
gives the ``best guess'' estimates
in the sense of having minimum variance (Hamilton 1997a; T97;
Bond, Jaffe \& Knox 2000), 
and also has the advantage of being independent of the number of 
bands used in the limit of high spectral resolution.
It was used for \fig{qFig} and \fig{Ffig}.
Here the window functions are simply the rows of the Fisher matrix,
and are seen to be rather broad.
All entries of $\F$ are guaranteed to be positive as proven in PTH01,
which means not only that all windows are positive (which is good)
but also that all measurements are positively correlated (which is bad).

Another interesting choice is (T97)
$\M=\F^{-1}$, which gives $\W=\I$.
In other words, all window functions are Kronecker
delta functions, and $\phat$ gives completely unbiased estimates 
of the band powers, with 
$\expec{\ph_i}=p_i$ regardless of what values the other band 
powers take. This gives an answer similar to the maximum-likelihood
method (THSVS98), and the covariance matrix of
\eq{pCovarEq} reduces to $\F^{-1}$.
A serious drawback of this choice is that that if we have
sampled the power spectrum on a scale finer than the inverse survey
size in an attempt to retain 
all information about wiggles {\etc}, this covariance matrix
tends to give substantially larger error bars 
($\Delta p_i\equiv\M_{ii}^{1/2}=[(\F^{-1})_{ii}]^{1/2}$)
than the first method, anti-correlated between neighboring bands.

The two above-mentioned choices for $\M$ both tend to 
produce correlations between the band power error bars.
The minimum-variance choice generally gives 
positive correlations, since the Fisher matrix cannot 
have negative elements, whereas 
the unbiased choice tends to give
anticorrelation between neighboring bands.
The choice (Tegmark \& Hamilton 1998; Hamilton \& Tegmark 2000) 
$\M=\F^{-1/2}$ with the rows renormalized
has the attractive property of making the errors uncorrelated,
with the covariance matrix of \eq{pCovarEq} 
diagonal. The corresponding window functions $\W$ are
plotted in \fig{Wfig}, and are seen to be 
quite well-behaved, even narrower than those in
\fig{Ffig} while remaining positive.\footnote{
The reader interested in mathematical challenges will be interested to know
that it remains a mystery to the authors 
why this $\F^{1/2}$ method works so
well. We have been unable to prove that $\F^{1/2}$ has no negative
elements (indeed, counterexamples can be contrived), yet the method
works like magic in practice in all LSS and CMB applications we
have tried.
}
This choice, which is the one we make in this paper, 
is a compromise between the two first ones:
it narrows the minimum variance window functions at the cost of
only a small noise increase, with uncorrelated noise as an extra bonus.
The minimum-variance band power estimators are essentially a smoothed version 
of the uncorrelated ones, and 
their lower variance was 
paid for by correlations which reduced the effective number of independent 
measurements.

\subsubsection{Disentangling the three power spectra}

The fact that we are measuring three power spectra rather than one
introduces an additional complication.
As illustrated by \fig{disentanglementFig}, an estimate 
of the power in one of the three spectra generally 
picks up unwanted contributions (``leakage'') from the other two,
making it complicated to interpret.
Although the above-mentioned $\F^{-1}$-method in principle eliminates
leakage completely, the cost in terms of increased error bars is found
to be prohibitive. We therefore follow HTP00 in adopting the following 
procedure for disentangling this three power spectra:
For each of the 49 $k$-bands, we take linear combinations
of the gg, gv and vv measurements such that the unwanted parts
of the window functions average to zero. 
This procedure is mathematically identical to that
used in Tegmark \& de Oliveira-Costa (2001) for separating different types
of CMB polarization, so the interested reader is referred there for the
explicit equations.
The procedure is 
illustrated in \fig{disentanglementFig}, and by construction has 
the property that leakage is completely eliminated if the 
true power has the same shape (not necessarily the same amplitude)
as the prior.
We find that this method works quite well 
(in the sense that the unwanted windows do not merely
average to zero) for the most accurately 
measured band powers, in particular the central parts of 
the gg-spectrum, with slightly poorer leakage elimination for 
bands with larger error bars.

\begin{figure} 
\vskip-0.8cm
\centerline{\epsfxsize=\figsize\epsffile{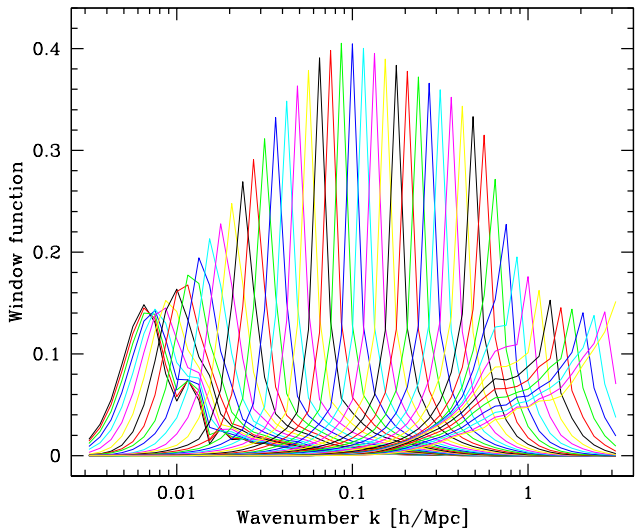}}
\vskip\smbotskip
\caption[1]{\label{Wfig}\footnotesize%
The window functions (rows of the $gg$-portion of $\W$) are shown
using decorrelated estimations.
The $\ith$ row of $\W$ typically peaks at the
$\ith$ band, the scale $k$ that the band power 
estimator $\ph_i$ was designed to probe.
Comparison with \fig{Ffig} shows that
decorrelation makes all windows substantially narrower.
}
\end{figure}

\begin{figure} 
\centerline{\epsfxsize=\figsize\epsffile{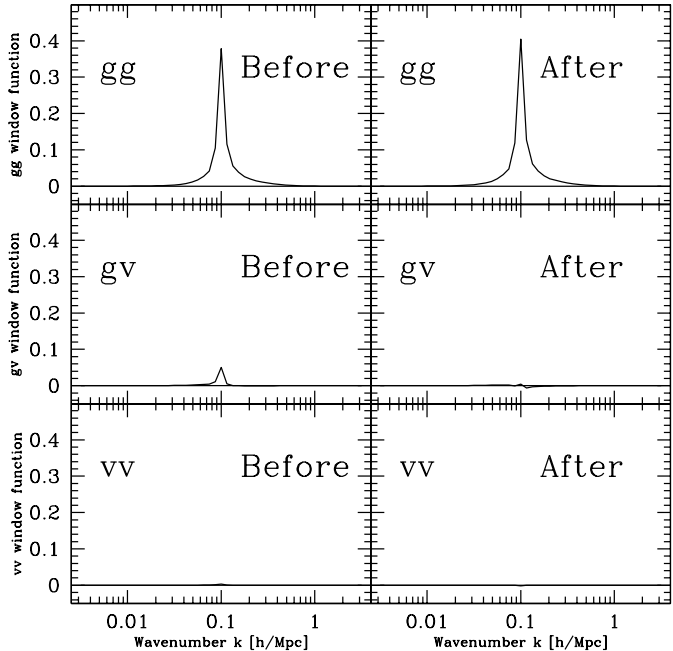}}
\caption[1]{\label{disentanglementFig}\footnotesize%
The window function for our measurement of the 25th
band of the galaxy-galaxy power is shown before
(left) and after (right) disentanglement.
Whereas unwanted leakage of gv and vv power 
is present initially, these unwanted
window functions both average to zero afterward.
The success of this method hinges on the fact that 
since the three initial functions (left) have similar shape,
it is possible to take linear combinations of
them that almost vanish (right). 
}
\end{figure}

The window functions plotted in \fig{Wfig} are the 
gg-windows after disentanglement.
Note that although our disentanglement introduces correlations
between the gg, gv and vv measurements for a given $k$-band, different
$k$-bands remain uncorrelated.

\begin{figure} 
\vskip-0.5cm
\epsfxsize=17cm\hglue-2mm\epsffile{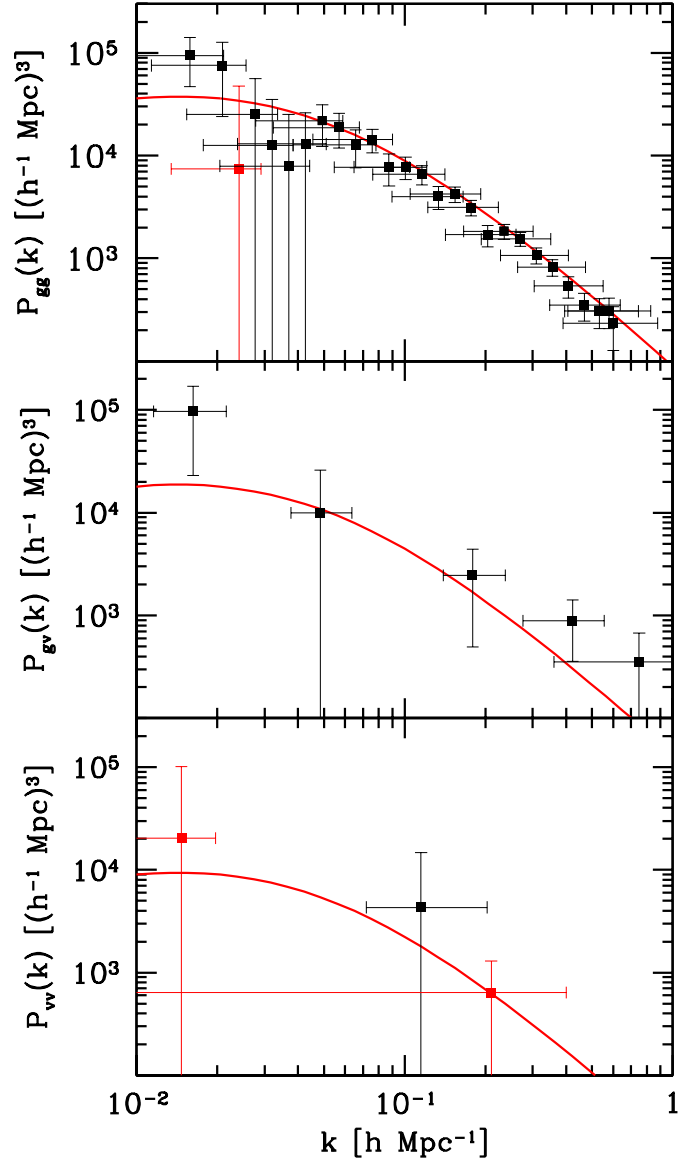}
\vskip-0.4cm
\caption[1]{\label{power_all3_binnedFig}\footnotesize%
Decorrelated and disentangled measurements of the galaxy-galaxy
power spectrum (top), the galaxy-velocity power spectrum (middle)
and the velocity-velocity) power spectrum (bottom) for the baseline 
galaxy sample.
Red points represent negative values --- since the points
are differences between two positive quantities 
(total power minus expected shot noise power), they can
be negative when the signal-to-noise is poor.
Each points is plotted at the $k$-value that is the 
median of its window function, and $68\%$ of this 
function is contained within the range of the horizontal bars. 
The curves shows our prior power spectrum.
Note that most of the information in the survey is on 
the galaxy-galaxy spectrum. 
Band-power measurements with very low information 
content have been binned into fewer (still uncorrelated) bands.

}
\end{figure}

\section{Results}
\label{ResultsSec}

\subsection{The three power spectra}

Our basic results are shown in \fig{power_all3_binnedFig}.
The single most striking feature of this plot is clearly that
the 2dFGRS is an amazing data set 
with unprecedented constraining power. 
The window functions in Figure~\ref{Wfig}
are seen to be quite narrow despite the complicated
survey geometry. The galaxy-galaxy power is constrained
to 20\% 
or better over an order of magnitude in length scale,
in about a
dozen uncorrelated bands centered around $k\sim 0.1\hperMpc$.
Whereas the increase in error bars on large scales 
reflects the finite survey volume, the lack of information on small
scales is caused by our analysis being limited to the first
4000 PKL-modes.
Comparing \fig{power_all3_binnedFig} with \fig{qFig} serves as a sobering reminder 
of the importance of decorrelating and disentangling the measurements to avoid 
a misleadingly rosy picture of how well one can do.

Whereas $\Pgg(k)$ is well measured, 
\fig{power_all3_binnedFig} shows that the information
about $\Pgv(k)$ is quite limited and that on $\Pvv(k)$ 
almost nonexistent. 
To avoid excessive cluttering in \fig{power_all3_binnedFig},
band-power measurements with very low information 
content have been binned into fewer (still uncorrelated) bands.
The main cause of these large error bars 
is that the information on $\Pvv$ and $\Pgg$ comes
from the quadrupole and hexadecapole moments of the
clustering anisotropy, which are intrinsically small and hence poorly 
constrained quantities.
However, the problem may be exacerbated by 
the lack
of large contiguous angular regions in the current data, 
impeding accurate comparisons of angular and radial 
clustering (the situation is similar for the SDSS; Zehavi {\etal} 2002),
and should improve as the survey nears completion and gets
more filled in. This effect is evident from a comparison with the 
results from the much more contiguous PSCz survey (HTP00):
the error bars on $\Pgg$ are appreciably larger for PSCz than 2dFGRS,
but those on redshift distortions (say $\beta$) are comparable.

In the remainder of this paper, we will address two separate
issues in turn: redshift-space distortions/biasing ($\beta$,$r$) 
and the detailed shape of the galaxy-galaxy power spectrum
(model fits, evidence for baryonic wiggles, \etc).

\subsection{Constraints on redshift space distortions}

\begin{figure} 
\centerline{\epsfxsize=\figsize\epsffile{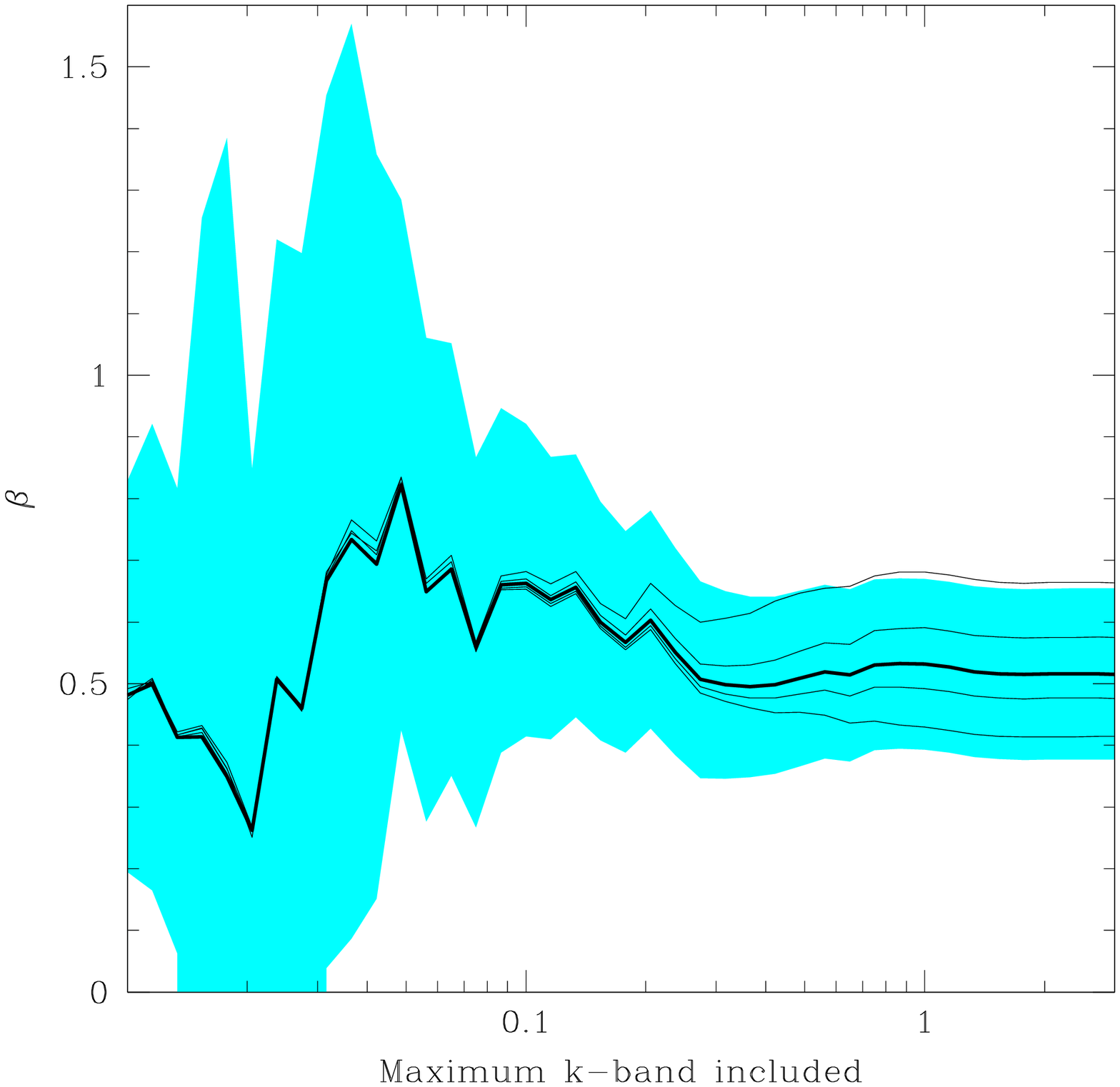}}
\caption[1]{\label{betaFig}\footnotesize%
The blue/grey band shows the $1\sigma$ allowed range
for $\beta$, assuming $r=1$ and the {\it shape} of the prior $\protect\Pgg(k)$
but marginalizing over the power spectrum normalization,
using FOG compression with density threshold $1{+}\delta_c=100$.
These fits are performed 
cumulatively, using all measurements for all wavenumbers $\le k$.
From bottom to top, the five curves show the best fit $\beta$ for
FOG thresholds $1{+}\delta_c=\infty$ (no FOG compression),
200, 100 (heavy), 50 and 25. 
}
\end{figure}

\begin{figure} 
\vskip-5.5cm
\centerline{\epsfxsize=\figsize\epsffile{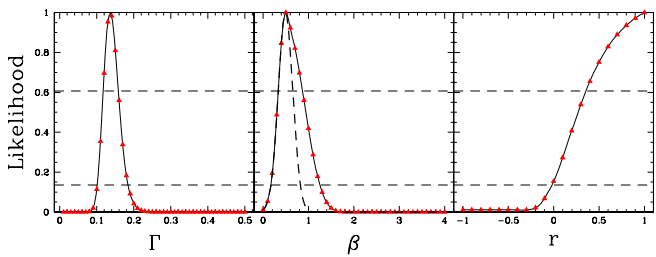}}
\vskip-0.5cm
\caption[1]{\label{gbrFig}\footnotesize%
1-dimensional likelihood curves for $\Gamma$, $\beta$ and $r$ 
are shown after marginalizing over the power spectrum 
normalization and the other parameters using our baseline
($1{+}\delta_c=100$) finger-of-god compression.
The  68\% and 95\% constraints are where the curves
intersect the dashed horizontal lines. The dashed curve 
in the middle panel shows how the $\beta$-constraints
tighten up when assuming $r=1$.
}
\end{figure}

\begin{figure} 
\centerline{\epsfxsize=\figsize\epsffile{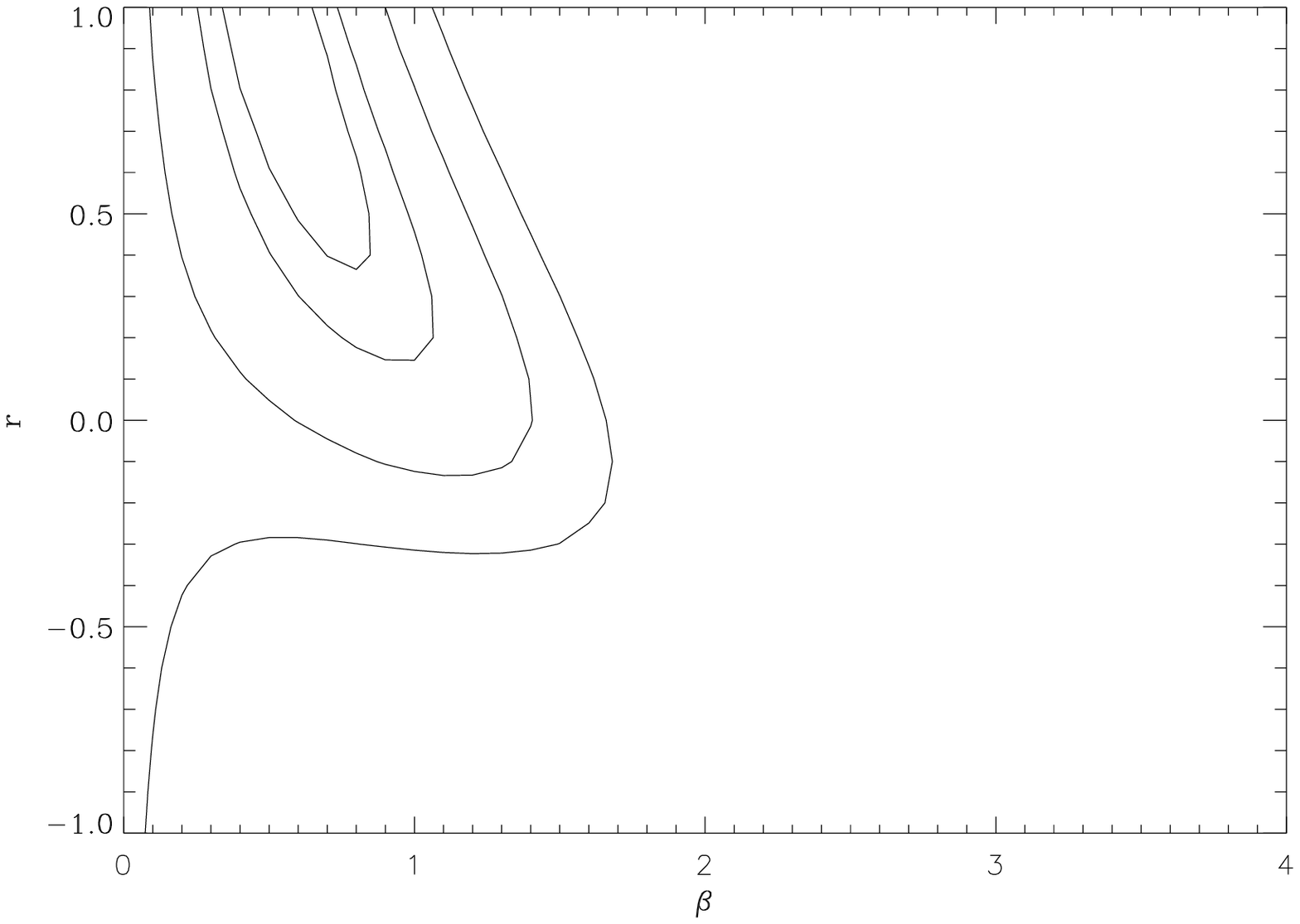}}
\caption[1]{\label{brFig}\footnotesize%
Constraints in the $(\beta,r)$ plane are shown
for our baseline
($1{+}\delta_c=100$) finger-of-god compression, using all measurements with
$k<0.3$h/Mpc and marginalizing over the power spectrum normalization
for fixed spectral shape.
The four contours correspond to  
$\Delta\chi^2=1$, $2.29$, $6.18$ and $11.83$, 
and would enclose  39\%,  68\%, 95\% and 99.8\% of the probability,
respectively, if the likelihood function were Gaussian.
}
\end{figure}

As seen from Figure~\ref{power_all3_binnedFig},
the constraints on $\Pgv(k)$ and $\Pvv(k)$ from 2dFGRS
are too weak to allow $\beta(k)$ and $r(k)$
to be measured reliably as a function of scale.
As data on Large Scale Structure improve,
it should become possible to accomplish such a measurement,
and thereby to establish quantitatively
the scale dependence of biasing at linear scales.
In the meantime we limit ourselves to the less ambitious goal
of measuring overall parameters $\beta$ and $r$, simply 
treating them as scale-independent constants.
This has not been previously done for the case of $r$.
Such scale-independence of bias on linear scales
is a feature of local bias models
(Coles 1993;
Fry and Gazta\~naga 1993;
Scherrer \& Weinberg 1998;
Coles, Melott \& Munshi 1999;
Heavens, Matarrese \& Verde 1999).

For our redshift-distortion analysis, we
employ a simple scale-invariant power spectrum $\Pgg(k)$ of the BBKS form
(Bardeen {\etal} 1986), parametrized by an amplitude $\sigma_8$ and 
a ``shape parameter'' $\Gamma$ that on a log plot
shifts the curve vertically and horizontally, respectively.
We will use more physically motivated 
power spectra with baryon wiggles {\etc} in \sec{CosmologySec} 
--- we tried various alternative parametrizations, and found that the
detailed form had essentially no effect on the $(r,\beta)$-constraints,
since they come from the ratios of the three spectra, not from their shapes.
Our model for the underlying band power vector $\p$ thus depends on 
four parameters $(\Gamma,\sigma_8,\beta,r)$.
We map out the likelihood function $L = e^{-\chi^2/2}$ 
using \eq{chi2Eq} on a fine grid in this parameter space,
and compute constraints on individual parameters by 
marginalizing over the other parameters as described
in Tegmark \& Zaldarriaga (2000), maximizing
rather than integrating over them.
The results are plotted in figures  
\ref{betaFig}, \ref{gbrFig} and \ref{brFig}.

\clearpage

\Fig{betaFig} assumes $\Gamma=0.14$, $r=1$ (the best fit
values) and explores how the results change as we include information
from smaller and smaller scales. As will be discussed 
in more detail in \sec{DiscussionSec}, non-linear effects 
invalidate the Kaiser approximation for redshift space distortions
on small scales. A smoking gun signature of such nonlinearities 
is $r$ and hence the best-fit $\beta$ dropping and ultimately going negative, 
as nonlinear ``fingers of god'' (FOGs) reverse
the effect of linear redshift distortions.
The fact that \fig{betaFig} does not show this effect is reassuring evidence
that little small-scale information is present in our data.
This is of course by design, since our PKL-modes contain 
contributions only from $\l\le 40$, corresponding to 
a comoving distance around $20\hMpc$ at the characteristic survey depth
of 400$\hMpc$.
This lack of small-scale information in our PKL-modes
is also reflected in the error bars on $\beta$, 
which are seen to stop decreasing around $k\sim 0.2\hperMpc$.

\Fig{betaFig} also shows how the results depend on the FOG removal
described in \sec{fogSec}.
The curves are seen to diverge markedly around $k\sim 0.2\hperMpc$,
with the FOG-related uncertainty becoming as large as the statistical
error bars for $k\sim 1\hperMpc$. We will return to these nonlinearity  
issues in \sec{nonlinearSec} below.

\Fig{gbrFig} shows the constraints on $\Gamma$, $\beta$ and $r$
after marginalizing over the other parameters.
The best fit model is 
$\Gamma=0.14$, $\beta=0.50$, $r=1$, $\sigma_8=0.99$.
The reason that the constraints on $\beta$ are so weak
is illustrated in \fig{brFig}: there is a degeneracy
with $r$. \Fig{power_all3_binnedFig} shows that our information about 
redshift distortions is coming predominantly from 
$\Pgv(k)$, not from the poorly constrained $\Pvv(k)$, 
so we are to first order measuring the combination $\beta r$
rather than $\beta$ and $r$ individually.
Imposing the prior $r=1$, as was implicitly done in 
Peacock {\etal} and almost all prior work, therefore
tightens the upper limit on $\beta$ substantially, 
as shown by the dashed curve in \fig{gbrFig}.

\subsection{The galaxy-galaxy power spectrum alone}

The previous subsection discussed the 2dFGRS constraints on redshift 
space distortions, essentially the ratios of the power spectra
$\Pgg(k)$, $\Pgv(k)$ and $\Pvv(k)$, without regard to their shape.
Let us now do the opposite, and focus on the shape of the
galaxy power spectrum $\Pgg(k)$.
The success of the disentanglement scheme illustrated in 
\fig{disentanglementFig} implies that 
the galaxy power spectrum plotted in 
\fig{power_all3_binnedFig} is robust,
essentially independent of what the 
power spectra $\Pgv(k)$ and $\Pvv(k)$ are doing.
However, this robustness came at a price in terms of increased error bars.
Assuming that all three power spectra have essentially the same shape,
but not the same amplitudes, we compute a more accurate
estimate of $\Pgg(k)$ as follows.

\begin{figure} 
\vskip-0.4cm
\centerline{\epsfxsize=\figsize\epsffile{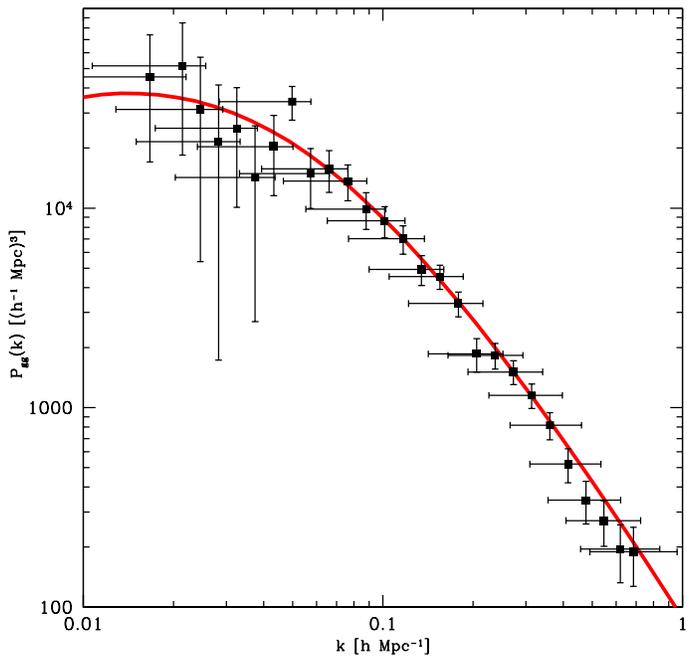}}
\caption[1]{\label{power_ggFig}\footnotesize%
The decorrelated galaxy-galaxy power spectrum is shown for the 
baseline galaxy sample
assuming $\beta=0.5$ and $r=1$. As discussed in the text, 
uncertainty in $\beta$ and $r$ contribute to an overall 
calibration uncertainty of order $12\%$ which is
not included in these error bars.
}
\end{figure}

We first assume some fixed values for $\beta$ and $r$.
This allows us to eliminate $\Pgv(k)$ and $\Pvv(k)$
using \eq{fzapEq}, reducing the size of our parameter
vector $\p$ from $3\times 49=147$ to 49 and our Fisher
matrix to size $49\times 49$, and gives
49 decorrelated estimators of $\Pgg(k)$.
The result assuming $\beta=0.5$, $r=1$ 
(our best fit values) is shown in 
\fig{power_ggFig}. We perform no binning here
except averaging the noisy bands with $k<0.02$ and $k>0.8$ into single bins 
to reduce clutter. 
We then repeat this exercise for a range of 
values of $\beta$ and $r$ consistent with
our analysis in the previous subsection to quantify 
the uncertainty these parameters introduce.
We find these uncertainties to be quite small,
as expected considering the small initial leakage 
of gv and vv power (see \fig{disentanglementFig}),
and can therefore quantify the added uncertainty 
$\delta\Pgg$ to first order as
\beq{derivativeEq}
\delta\ln\Pgg(k) 
= \left|{\partial\ln\Pgg(k)\over\partial(\beta r)}\right|\delta(\beta r)
+ \left|{\partial\ln\Pgg(k)\over\partial(\beta^2)}\right|\delta(\beta^2).
\eeq
Numerically, we find these two derivatives to 
be approximately $-0.2$ and $-0.04$, respectively,
essentially independent of $k$.
This scale-independence is not surprising in the
small-angle limit, where these derivatives would involve
simply various average moments of $\mu$, the angle between the
$\k$-vector and the line of sight.
Assuming uncertainties $\delta\beta=0.15$ and
and $\delta r=0.5$, \eq{derivativeEq} thus gives
$\delta\ln\Pgg(k)\approx 0.12$, the second term
being negligible relative to the first.
In conclusion, the uncertainties in \fig{power_ggFig}
induced by uncertainties about $\beta$ and $r$ can be 
summarized as simply an overall multiplicative calibration error of order 12\%
for the measured power spectrum.

\section{How reliable are our results?}
\label{SystematicsSec}

How reliable are the results presented in the previous section? 
In this section, we perform a series of tests, 
both of our software and algorithms
and of potential systematic errors. We also discuss the 
underlying assumptions that are likely to be most
important for interpreting the results.

\subsection{Validation of method and software}
\label{MonteSec}

Since our analysis consists of a number of numerically non-trivial steps,
it is important to test both the software and the underlying methods.
We do this by generating $\nmonte=100$ Monte Carlo simulations of the
2dFGRS catalog with a known power spectrum, processing them through our
analysis pipeline and checking whether they give the correct answer
on average and with a scatter corresponding to the predicted error bars.
We found this end-to-end testing to be quite useful in all phases of this 
project --- indeed, we had to run the pipeline 43 times until everything 
finally worked...

\subsubsection{The mock survey generator}

Standard N-body simulations would not suffice for our precision test,
because of a slight catch-22 situation: the true non-linear power spectrum
of which an N-body simulation is a realization (with shot noise added)
is not known analytically, and is usually estimated by measuring it
from the simulation --- but this is precisely the step that we wish
to test.
We therefore generate realizations that are firmly in the linear regime,
returning to nonlinearity issues below.
We do this as described in PTH01, with
a test power spectrum of the simple
Gaussian form $P(k)\propto e^{-(Rk)^2/2}$ with
$R=32\,h^{-1}\Mpc$, normalized so that the rms fluctuations
$\expec{\delta^2}^{1/2}=0.2$.

\subsubsection{Testing the PKL pixelization}

\begin{figure} 
\vskip\smtopskip
\centerline{\epsfxsize=\figsize\epsffile{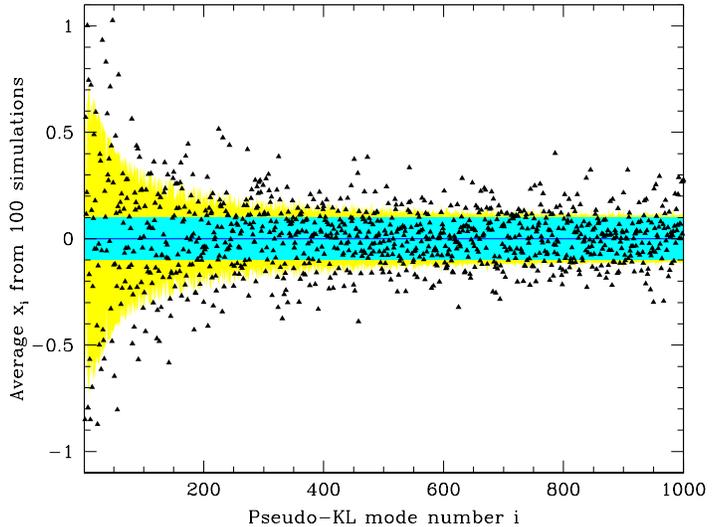}}
\vskip\smbotskip
\caption[1]{\label{xmonte1Fig}\footnotesize%
The triangles show the elements $x_i$ of the data vector $\x$
(the pseudo-KL expansion coefficients) averaged
over 100 Monte-Carlo simulations of the 
baseline galaxy sample.
If the algorithms and software are correct, then 
their mean should be zero and about
$68\%$ of them should lie within
the shaded yellow/grey region giving their standard deviation.
}
\end{figure}

\Fig{xmonte1Fig} 
shows the result of processing the Monte Carlo simulations
through the first step of the analysis pipeline, \ie, 
computing the corresponding Pseudo-KL expansion coefficients
$x_i$. This is a sensitive test of the mean correction 
given by \eq{meanDefEq2}, which can be a couple of orders of magnitude
larger than the scatter in \fig{xmonte1Fig} for some modes.  
A number of problems with the 
radial selection function integration and the spherical harmonic expansion
of the angular mask in our code were discovered in this way.
After fixing these problems, the coefficients $x_i$ became consistent
with having zero mean as seen in the figure.

\begin{figure} 
\vskip\smtopskip
\centerline{\epsfxsize=\figsize\epsffile{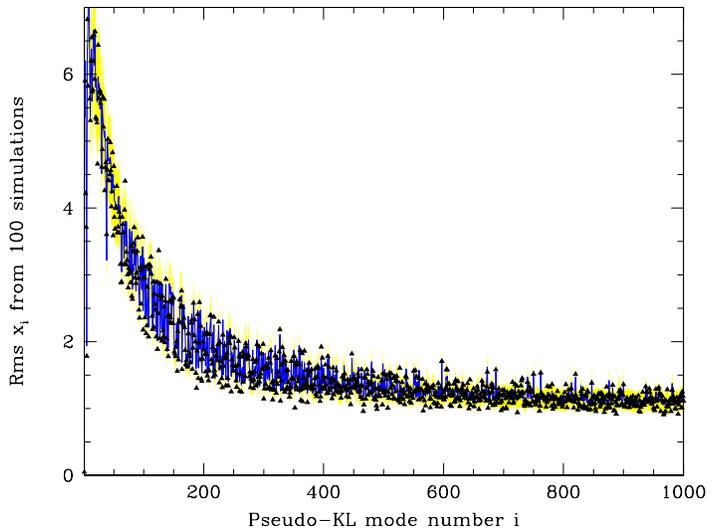}}
\vskip\smbotskip
\caption[1]{\label{xmonte2Fig}\footnotesize%
The triangles show the rms fluctuations of the elements $x_i$ 
from 100 Monte-Carlo simulations.
If the algorithms and software are correct, then 
the expectation value of this rms is given by the thin blue curve,
and most of them should scatter in the yellow/grey region.
}
\end{figure}

\begin{figure} 
\centerline{\epsfxsize=\figsize\epsffile{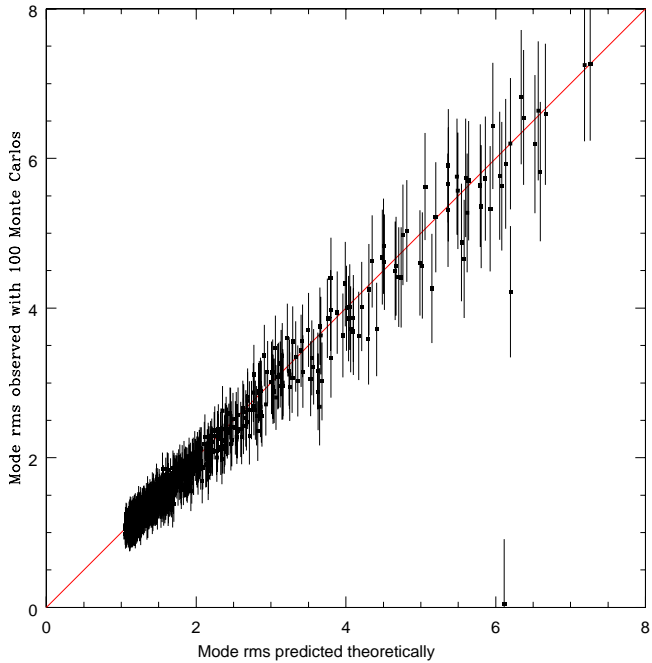}}
\caption[1]{\label{xmonte3Fig}\footnotesize%
In this alternative representation of the test from 
\fig{xmonte2Fig}, most of the vertical lines
should intersect the $45^\circ$ line
if the algorithms and software are correct.
}
\end{figure}

\Fig{xmonte1Fig} 
also shows that the scatter in the modes
is consistent with the predicted standard deviation 
$\sigma_i=(\C_{ii}/\nmonte)^{1/2}$ (shaded region), with most of the
the fluctuations being localized to modes probing large scales
(with $i$ being small).
A more sensitive test of this scatter is shown in
\Fig{xmonte2Fig}, which shows that the theoretically predicted
variance for each mode agrees with what is observed in the
100 Monte Carlo realizations.
Since crowding makes it hard to verify all modes in
this plot, an alternative representation of this test is
shown in \fig{xmonte3Fig}.

\begin{figure} 
\vskip\smtopskip
\centerline{\epsfxsize=\figsize\epsffile{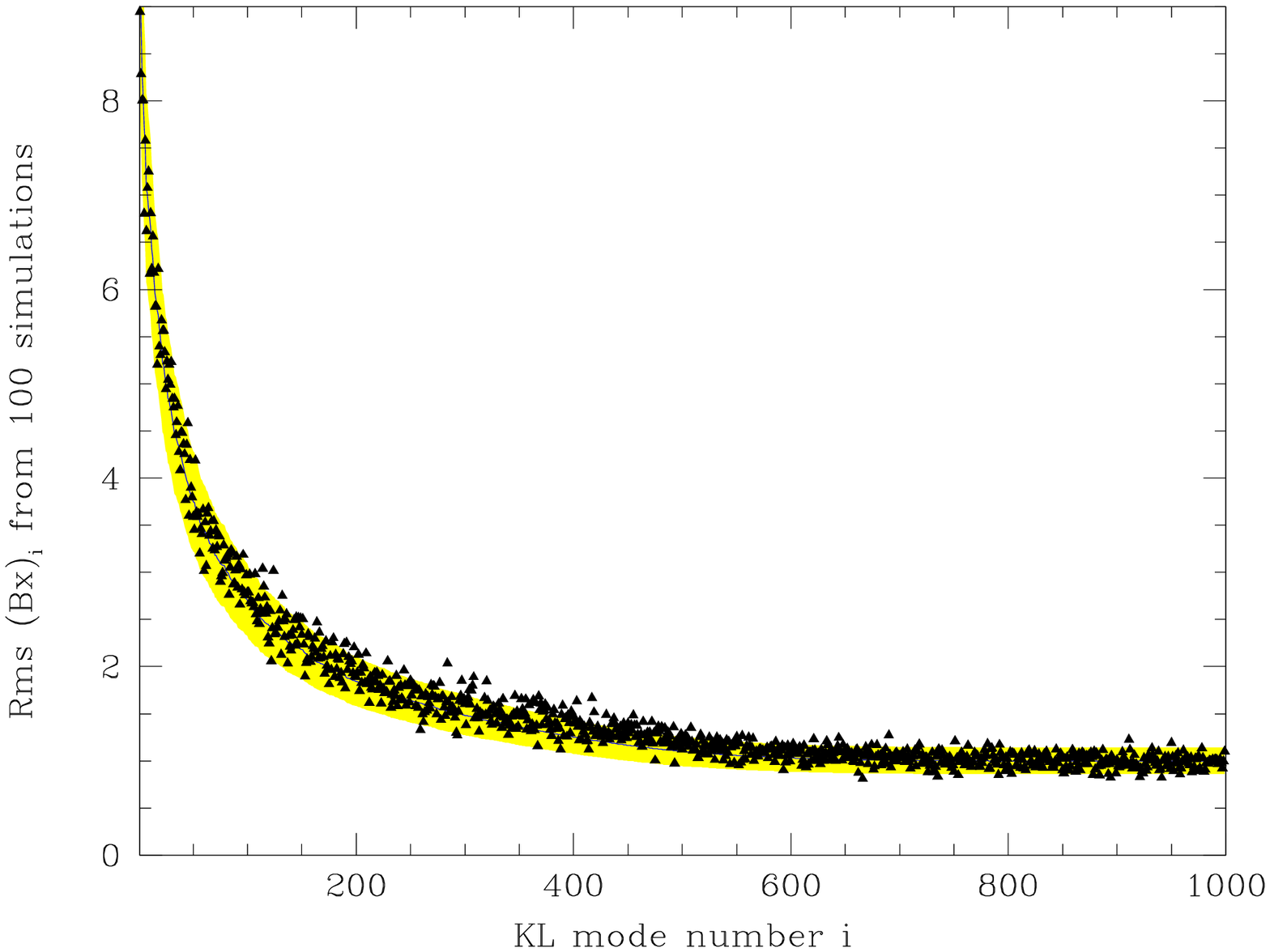}}
\vskip\smbotskip
\caption[1]{\label{WxmonteFig}\footnotesize%
The triangles show the rms fluctuations of the elements $(\B\x)_i$ 
from 100 Monte-Carlo simulations.
If the algorithms and software are correct, then 
the expectation value of this rms is given by the thin blue curve,
and most of them should scatter in the yellow/grey banana-shaped 
region.
}
\end{figure}

Although these tests verify that the mean and variance of each mode come out 
as they should, they are not sensitive to errors in the off-diagonal
elements of the covariance matrix $\C$, \ie, to incorrect correlations
between the mode coefficients. 
To close this loophole, \fig{WxmonteFig} shows the 
scatter in the true KL-modes ($\y=\B\x$),
illustrating agreement with the theoretical variance prediction
even in this alternative basis where all coefficients $y_i$
should be uncorrelated.
Note that the expected variance decreases monotonically here,
as opposed to in \fig{xmonte2Fig}, since the true KL-modes are
strictly sorted by decreasing variance.

\subsubsection{Testing the quadratic compression, Fisher decorrelation
and disentanglement}

\begin{figure} 
\centerline{\epsfxsize=\figsize\epsffile{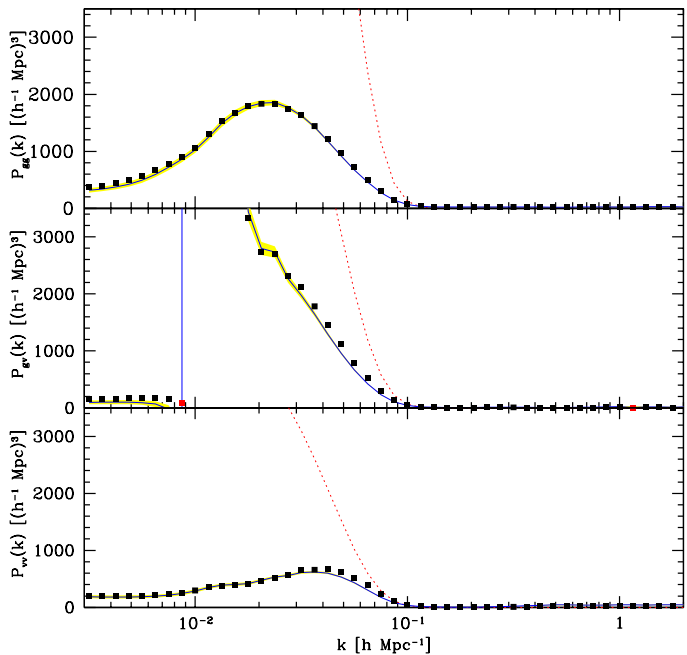}}
\caption[1]{\label{pmonte1Fig}\footnotesize%
The triangles show the decorrelated and disentangled band-power 
estimates $\ph_i$, averaged
over 100 Monte-Carlo simulations of the baseline galaxy sample.
If the algorithms and software are correct, then
this should recover the window-convolved input power spectrum
$\W\p$, plotted as a thin blue line.
The thin shaded yellow/grey band indicates the expected scatter.
The harmless discontinuity in the middle panel is an artifact of the
disentangled galaxy-velocity windows having negative area on the largest
scales where there is essentially no information available.
}
\end{figure}

\begin{figure} 
\centerline{\epsfxsize=\figsize\epsffile{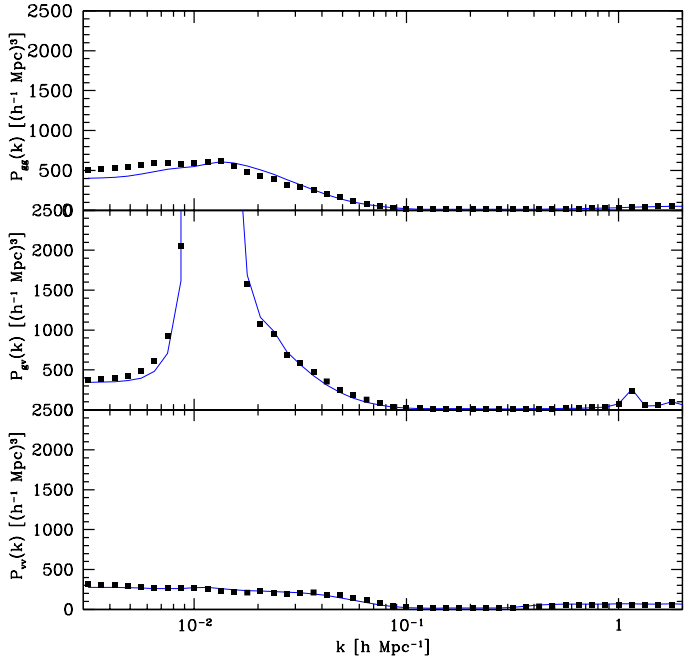}}
\caption[1]{\label{pmonte2Fig}\footnotesize%
Same as the previous figure, but testing 
the error bars $\Delta p_i$ rather than the power itself.
The triangles show the observed rms of the power spectrum estimates
from 100 simulations and the solid blue curve shows the predicted 
curve around which they should scatter.
}
\end{figure}

Figures~\ref{pmonte1Fig} and~\ref{pmonte2Fig} show 
the result of processing the Monte Carlo simulations
through the remaining steps of the analysis pipeline, \ie, 
computing the raw quadratic estimator vector $\q$ and, from it,
the decorrelated and disentangled
band-power vector $\phat$.
The mean recovered power spectra are seen to be in excellent agreement
with the Gaussian prior used in the simulations (\fig{pmonte1Fig})
convolved with the window functions,
and the observed scatter is seen to be consistent
with the predicted error bars (\fig{pmonte2Fig}).
These two figures therefore constitute an end-to-end test
of our data analysis pipeline, since errors in any of the
many intermediate steps would have shown up here at some level.
Since information from large numbers of modes contributes to each 
$p_i$, the scatter is seen to be small. Therefore,
even quite subtle bugs and inaccuracies can be (and were!)
discovered and remedied as a result of this test.

\subsection{Robustness to method details}

Our analysis pipeline has a few ``knobs'' that can be
set in more than one way. This section discusses the 
sensitivity to such settings.

\subsubsection{Effect of changing the prior}
\label{PriorSec}

The analysis method employed assumes a ``prior'' power
spectrum via \eq{CsumEq}, both to compute band power error bars
and to find the galaxy pair weighting that minimizes them.
As mentioned, an iterative approach was adopted starting
with a simple BBKS model, then 
shifting it vertically and horizontally to better fit the resulting measurements and 
recomputing the measurements a second time.
To what extent does this choice of prior affect the results?
On purely theoretical grounds
(\eg, Tegmark, Taylor \& Heavens 1997), one expects a grossly incorrect
prior to give unbiased results but with unnecessarily large
variance. If the prior is too high, the sample-variance 
contribution to error bars will be overestimated and vice versa.
This hypothesis has been extensively 
tested and confirmed in the context of power spectrum measurements
from both the Cosmic Microwave Background 
(\eg, Bunn 1995) and galaxy redshift surveys
(PTH01), confirming that the correct result is recovered on average even when using 
a grossly incorrect prior. 
In our case, the prior by construction agrees quite well with the actual
measurements (see \fig{power_all3_binnedFig}), so the quoted error 
bars should be reliable as well.

\subsubsection{Effect of changing the number of PKL modes}

We have limited our analysis to the first $N=4000$ PKL modes
whose angular part is spanned by spherical harmonics with 
$\l\le 40$. This choice was a tradeoff between the desire to
capture as much information as possible about the galaxy survey
and the need to stay away from small scales where non-linear effects
invalidate the Kaiser approximation to redshift distortions. 
To quantify our sensitivity to these choices, we repeated the
entire analysis using 500, 1000, 2000 and 4000 modes.
Our power spectrum measurements on the very largest scales were
recovered even with merely 500 modes.
As we added more and more modes (more and mode small-scale information), 
the power measurements converged to those in \fig{power_all3_binnedFig}
for larger and larger $k$.
The rising part of the envelope in \fig{Ffig} remained essentially unchanged, 
merely continuing to grow further as more modes were added, so the 
turnover of this envelope directly shows the $k$-scale beyond
which we start running out of information.
The version of \Fig{Ffig} shown in this paper 
indicates that our 4000 PKL modes
have captured essentially all cosmological information from
the 2dFGRS for $k\simlt 0.1$.

\subsubsection{Numerical issues}

\begin{figure} 
\vskip\smtopskip
\centerline{\epsfxsize=\figsize\epsffile{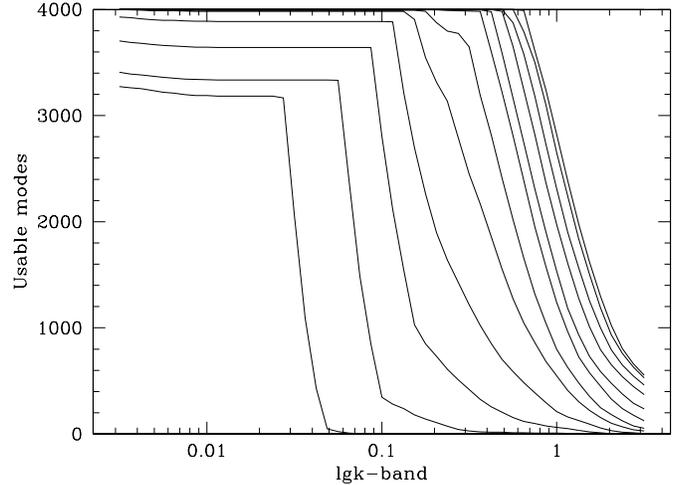}}
\vskip\smbotskip
\caption[1]{\label{usableFig}\footnotesize%
Numerical convergence. The figure shows for how many of our 4000 PKL modes
the numerical calculations are converged to accurately measure the power
up to a given wavenumber $k$.
From left to right, the 12 curves correspond to truncation at 
$\lcut=$20, 40, 60, 80, 100, 120, 140, 160, 180, 200, 220 and 240.
}
\end{figure}

The computation of the matrices $\P_i$ involves a summation over
multipoles $\l$ that should, strictly speaking, run from $\l=0$ to $\l=\infty$,
since the angular mask itself has sharp edges involving harmonics
to $\l = \infty$.
In practice, this summation must of course be truncated at some 
finite multipole $\lcut$. To quantity the effect of this truncation, we plot 
the diagonal elements of the $\P$-matrices as a function of $\lcut$
and study how they converge as $\lcut$ increases.
We define a given PKL-mode as having converged by
some multipole if subsequent $\l$-values contribute 
less than 1\% of its variance. \Fig{usableFig} plots the number
of usable PKL-modes as a function of wavenumber $k$,
defining a mode to be usable for our analysis only if it
is converged for all smaller wavenumbers $k'<k$ for 
all three power flavors ($\Pgg$, $\Pgv$ and $\Pvv$).
We use $\lcut=260$ in our final analysis, since this guarantees
that all 4000 modes are usable
for wavenumbers $k$ in the range $0-0.5\hperMpc$, \ie, 
comfortably beyond the large scales $0-0.3\hperMpc$ that are the
focus of this paper.  
With this cutoff, the computation of the $P$-matrices (which scales as
$\lcut^2$ asymptotically),
took about a week on a SunBlade1000 workstation.
Our power spectrum estimates are likely to remain fairly accurate 
as far out as we plot them, \ie, to $k\sim 1\hperMpc$, since \fig{usableFig} shows most modes remaining
usable out to this scale, and since we find that even the ones that do not 
meet our strict 1\% convergence criterion at every single band are generally
fairly accurately treated. Indeed, we repeated our entire analysis with 
$\lcut=120$ and obtained almost indistinguishable power spectra.

\subsection{Tests for problems with data modeling}

In \sec{DataSec}, we performed detailed modeling of the 
way in which the 2dFGRS data was selected, and produced a
uniform galaxy sample fully characterized by 
a selection function $\nbar(\r)$ of the separable 
form of \eq{SeparabilityEq}.
Let us now assess how sensitive our results are to potential
mis-estimates of $\nbar$, both angularly and radially, 
by discarding purely angular and radial modes from our analysis.

\begin{figure} 
\centerline{\epsfxsize=\figsize\epsffile{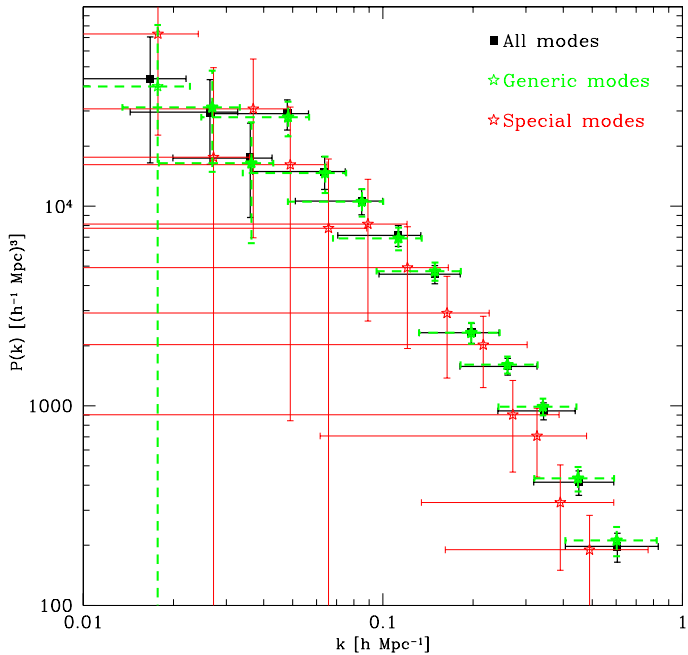}}
\caption[1]{\label{P3fig}\footnotesize%
Constraints on excess power in special modes.
Our 2dF power spectrum measurements from \fig{power_ggFig}
are averaged into fewer bands and compared with measurements 
using only special (radial, angular and local group) modes
and only generic (the remaining) modes (dashed).}
\end{figure}

\subsubsection{Robustness to angular problems}
\label{angularSec}

Angular modulations caused by dust extinction tend to have a
power spectrum rising sharply toward the largest scales
(Vogeley 1998), and is therefore of particular concern for the interpretation
of our leftmost bandpower estimates. 
The galaxy magnitudes are extinction corrected by the 2dFGRS team, 
using extinction map produced by Schlegel, Finkbeiner \& Davis
(1998), so any inaccuracies in this extinction model would masquerade as
excess large-scale power.
Inaccuracies in zero-point offsets or in the 
magnitude dependent completeness correction that we 
applied in \sec{muSec} could also introduce spurious angular power.

Of our 4000 modes, 147 are purely angular 
(see \fig{slicemodesFig} for an example), and as described in
\sec{PKLsec}, the remaining 3853 are orthogonal to them. 
This means that to first order, angular problems affect only these
147 PKL-coefficients $x_i$. We repeated our entire analysis with these 
coefficients discarded, and found that the error bars became 
so large for $k\simlt 0.03\hperMpc$ that no signal could be detected 
there. In other words, the information on the power spectrum 
on the very largest scales comes mainly from the purely angular modes.
On smaller scales, the measured power spectrum remained essentially unchanged.
Although we have no indication that angular problems are actually present,
it may be prudent to follow the 2dFGRS team and discard the information
on the very largest scales --- to be conservative, we therefore use only the
measurements for $k\ge 0.01\hperMpc$ to be conservative in
our likelihood analyses (for $\beta$, $r$ and cosmological parameters).

\subsubsection{Robustness to problems with the radial selection function}
\label{radialSec}

45 of our 4000 modes are purely radial (see \fig{slicemodesFig} for an example),
and are to first order the only ones affected by mis-estimates of
the radial selection function $\nbar(r)$.
Since accurate $k$-corrections and evolution modeling are notoriously challenging
to perform, we repeated our entire analysis with these 45 modes omitted
as a precaution. This resulted in a slight increase in error bars
on the largest scales, but much less noticeable than when we removed the 
angular modes as described above. This can be readily understood geometrically.
If we count the number of modes that probe mainly scales $k<k_*$,
then the number of purely radial, purely angular and arbitrary modes
will grow as $k_*$, $k_*^2$ and $k_*^3$, respectively.
This means that ``special'' modes (radial and angular) will make up a larger fraction
of the total pool on large scales (at small $k$), and that the purely 
radial ones will be outnumbered by the purely angular ones.

Percival {\etal} (2001) report that slight changes in $\nbar(r)$ did not
have a strong effect on the recovered 2dFGRS power spectrum,
and we confirm this. We repeated our analysis with a number of different
radial selection functions $\nbar(r)$, including the one from Colless {\etal} (2001) 
(the dashed curve in \fig{zhistFig}), finding only changes smaller than the error bars for 
$P(k)$ on the largest scales and no noticeable changes for larger $k$.

A final end-to-end test for problems with any special 
(angular, radial, or local group) modes is shown in \fig{P3fig}.
Here we have repeated the entire analysis twice, 
once excluding all the special modes and once using 
{\it only} the special modes (except the monopole). The latter is seen to 
give quite large error bars since only 196 modes are used
(4 local group,   147 angular and 45 radial), 
but all three are seen to be reassuringly consistent. In contrast, 
systematic problems with any special modes would tend to
add power to the special modes.
This shows that any misestimates of special modes is having a negligible impact on our
final results.

\subsection{Non-linearity issues}
\label{nonlinearSec}

A key assumption (essentially the only one) 
underlying our analysis is that
the Kaiser (1987) linear perturbation theory approach to 
redshift space distortions is valid.
This approximation is known to break down on small scales where nonlinear effects 
become important, which is why we have limited our analysis to large scales.

To be more precise, our basic measurement of 
$\Pgg(k)$, $\Pgv(k)$ and $\Pvv(k)$ assumes nothing at all, 
and measures the quantities that reduce
to the monopole, quadrupole and hexadecapole of power in the
in small-angle approximation (Hamilton 1998). However,
relating these three measured functions to 
$\beta(k)$ and $r(k)$ via \eq{fzapEq} does require 
the Kaiser approximation to be valid.

Substantial progress has recently been made in quantifying 
nonlinear effects on redshift distortions, using both perturbation
theory, gravitational $N$-body simulations and
semianalytic galaxy formation theory
(Hatton \& Cole 1997, 1999;
Scoccimarro {\etal} 1999;
Heavens, Matarrese \& Verde 1999;
Scoccimarro, Zaldarriaga \& Hui 1999; Hamilton 2000;
Seljak 2001; Scoccimarro \& Sheth in preparation).
The consensus is that nonlinear effects may be important 
even on scales as large as $k\sim 0.1-0.3\hperMpc$), although the critical scale
is sensitive to the type of galaxies involved via their bias properties
(Seljak 2001). Moreover, a generic smoking-gun signature of nonlinear effects is found to
be that the ratio $\Pgv(k)/\Pgg(k)$ starts dropping and eventually becomes negative, 
as nonlinear fingers-of-god reverse the signature of linear infall.
The ratio $\Pvv(k)/\Pgg(k)$ increases sharply in this regime.

Ideally, to do full justice to the 2dFGRS data set, one would like to 
perform a suite of nonlinear simulations until a realistic biasing scheme is 
found that reproduces all observed characteristics of the data.
The fast PTHalos approach (Scoccimarro \& Sheth 2002) suggests 
that such an ambitious approach may ultimately be feasible.
In the interim, the results obtained with analytic approximations 
must be interpreted with great caution.
Peacock {\etal} (2001) use the widespread approach of adding 
a nuisance parameter to the Kaiser formula, interpreted as a 
small-scale velocity dispersion (cite), and marginalizing over it.
This gives $\beta=0.43\pm 0.07$ from 141{,}000 2dFGRS
galaxies. 
Hatton \& Cole (1999) and Scoccimarro \& Sheth (in preparation) argue that this is approximation is inaccurate, underestimating the nonlinear corrections 
(hence underestimating $\beta$)
on large scales, and that the approximation of Hatton \& Cole (1999) is preferable. 

Given these important uncertainties, we adopt a more empirical approach, 
using the above-mentioned
$\Pgv$-drop in the data as an indicator of where to stop trusting the results.
This was also done in the PSCz analysis of Hamilton {\etal} (2001),
where $\beta$ was found to start dropping for $k\simgt 0.3\hperMpc$.
\Fig{power_all3_binnedFig} shows no indication of $\Pgv(k)/\Pgg(k)$
(basically the quadrupole-to-monopole ratio) dropping,
suggesting that our linear approximation is not 
seriously biasing our results on the large scales probed by
our PKL modes (which recover information fully down to $k\sim 0.1$ 
as described above).

To quantify further the effect of non-linearities empirically, 
we performed our entire analysis five times with different levels 
of finger-of-god (FOG) compression as described in \sec{fogSec}.
The five curves in \fig{betaFig} correspond to progressively more
aggressive compression with overdensity cutoffs
$1{+}\delta_c=\infty$, 200, 100, 50 and 25, respectively.
This corresponds to  6677, 7820, 8643 and 9124 FOG's compressed, involving 
18544, 24031, 29807 and 36098 galaxies, respectively.
\Fig{betaFig} shows that more aggressive FOG-compression has an effect with
the expected sign, increasing the best-fit $\beta$-value for $k\simgt 0.1$, 
and that the effect is reassuringly small compared to the statistical error bars.
Since a cluster is expected to have an overdensity around 200 when
it virializes, more later since the background density drops, 
thresholds $1{+}\delta_c<100$ are likely to be overkill --- we included 
the cases $1{+}\delta_c=50$ and $25$
in the figure merely to explore an extreme range of remedies.
By removing essentially all structures that are elongated along the 
line of sight, one of course creates an artificial excess of flattened 
structures, leading to an overestimate of $\beta$.
In conclusion, we believe that our estimate $\beta=0.49\pm 0.16$ is not
severely affected by nonlinearities. 
A conservative approach would be
to take our measurement without FOG compression and use 
it merely as a lower limit, giving 
$\beta > 0.26$ at $90\%$ confidence.
 
Non-linearities affect our analysis in a different way as well, leading to
slight underestimates of error bars. 
Our power spectrum measurements are simply certain second moments of the
data, and remain valid regardless of whether the underlying density field 
is Gaussian or not. The power spectrum variance, however, involves
fourth moments, and we have 
computed our error bars by making the Gaussian approximation 
to calculate these moments.
The standard rule of thumb is that this approximation 
underestimates the error bars on the correlation function $\xi(r)$
by a factor $[1+\xi(r)]^{1/2}$.
Norberg {\etal} (2001)  fit the 2dFGRS correlation function
to a power law $\xi(r)=(r/r_*)^{-\gamma}$ with 
correlation length $r_*=4.9\hMpc$ and 
slope $\gamma=1.71$. Taking $k\sim\pi/r$, 
this gives error bar correction factors 
$[1+(r_*k/\upi)^\gamma)]^{1/2}\approx 2\%$, $7\%$ and $13\%$
at $0.1$, $0.2$ and $0.3\hperMpc$, respectively. 
Here $\xi(r)$ should refer to the correlation function of the matter,
not of the galaxies, so if the 2dFGRS galaxies are biased with
$b>1$, the correction factors will be smaller.
In conclusion, although nonlinear error bar corrections certainly become 
important on very small scales, they are likely to be of only minor
importance on the large scales $k<0.3\hperMpc$ that are the focus of 
this paper.

\subsection{Bias issues}

Although our basic measurement of $\Pgg(k)$, $\Pgv(k)$ and $\Pvv(k)$ assumes 
nothing about biasing, a bias model is obviously necessary before the
results can be used to constrain cosmological models. We therefore comment
briefly on the bias issue here.

Substantially larger data sets such as the complete SDSS catalog
hold the promise of measuring 
$\beta(k)$ and $r(k)$ with sufficient accuracy to quantity 
their scale-dependence, if any.
\fig{power_all3_binnedFig} shows that
our present sample is still not quite large enough to place strong constraints
of this type. 

An alternate route to constraining $b(k)$ involves comparing 
the clustering amplitudes of various subsamples, selected by, say, 
luminosity or spectral type. Such comparisons can also 
constrain $r$ directly (Tegmark \& Bromley 1999; Blanton 2000).
It has been long known
that bright elliptical galaxies are
more clustered than
spirals, 
presumably because the former are more likely to reside in
clusters.
Recent subsample analysis of the 2dFGRS 
(Norberg {\etal} 2001a) 
and SDSS (Zehavi {\etal} 2002) have confirmed and 
further quantified this effect.

Since recent cosmological parameter analyses using $P(k)$-measurements
(most recently Wang {\etal} 2002 and Efstathiou {\etal} 2002)
have assumed that the
bias factor $b$ is scale-independent on linear scales, it is important
to note  that slight scale-dependence of bias is likely to be present 
in $\Pgg(k)$-measurements from a heterogeneous galaxy 
sample such as the 2dFGRS.
Most of the information about $\Pgg(k)$ on large scales comes 
from distant parts of the survey, where bright ellipticals
are over-represented since dimmer galaxies get excluded by
the faint magnitude limit. This could cause $b(k)$ to rise 
as $k\to 0$. 
If uncorrected, this effect could masquerade as evidence
for a redder power spectrum, \ie, one with a smaller 
spectral index $n$.

\Fig{power_ggFig} indeed suggests slightly more 2dFGRS
power on the largest scales than currently favored cosmological models
with constant bias would suggest, although this excess may also 
be caused by the angular or radial issues mentioned above.
Detailed power spectrum analysis of subsamples should settle
this issue.

\section{Discussion and conclusions}
\label{ConcSec}
\label{DiscussionSec}

To place our results in context, we will now briefly discuss how they compare with
other recent power spectrum measurements and with 
cosmological models.

\subsection{Comparison with other surveys}

\begin{figure} 
\centerline{\epsfxsize=\figsize\epsffile{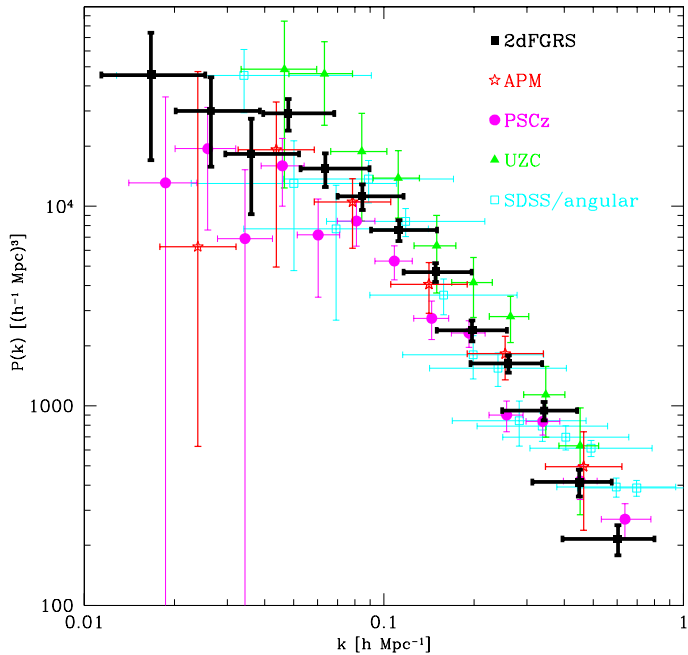}}
\caption[1]{\label{P1fig}\footnotesize%
Comparison with other power spectrum measurements.
Our 2dF power spectrum measurements from \fig{power_ggFig}
are averaged into fewer bands and compared with measurements 
from the PSCz (HTP00) and UZC (this work)  
redshift surveys as well as angular clustering in the 
APM survey (Efstathiou \& Moody 2001)
and the SDSS (the points are from Tegmark {\protect\etal} 2002
for galaxies in the magnitude range $21<r'<22$  --- see also 
Dodelson {\protect\etal} 2002).
}
\end{figure}

\Fig{P1fig} compares our 2dFGRS  power spectrum measurements from \fig{power_ggFig} (averaged into fewer bands to reduce clutter) with
measurements from other recent surveys.
The PSCz and UCZ redshift surveys were analyzed with 
the same basic method that we have employed here\footnote{
Since the UZC analysis in PTH01 did not include redshift space distortions,
we performed a complete reanalysis of that data set for this figure, 
expanding the 13342 galaxies surviving the cuts described in PTH01 
in 1000 PKL modes.
}, so a direct comparison involves no method-related interpretational issues. 
The 2dFGRS sample is seen to be slightly more biased
than PSCz, but slightly less biased than UZC. 
\Fig{P1fig} also suggests that 2dFGRS may have a slightly redder
power spectrum than PSCz.  This would also be consistent with the
scale-dependent bias scenario mentioned above --- the PSCz survey
would probably be less afflicted than 2dFGRS,
since the {\it IRAS}-selected galaxies in PSCz tend to avoid clusters.

Although the 2dFGRS error bars
are seen to be small compared the PSCz and UZC ones,
due to the larger sample size and survey volume, the
horizontal bars show that the 2dFGRS window functions are somewhat
broader. This is easy to understand: whereas PSCz and UZC cover
large contiguous sky regions, the 2dFGRS sky coverage is currently fragmented into 
a multitude of regions of small angular extent, exacerbating aliasing problems.
Indeed, since the characteristic width of 2dFGRS patches in the narrowest direction is 
more than an order of magnitude smaller than for PSCz or UZC
(of order $2^\circ$ rather than $\sim 60^\circ$), the fact that the windows are
only 2-3 times wider reflects the quality of the 2dFGRS survey design and 
the power of the quadratic estimator method.

The remaining two power spectra are interesting since they
were measured without use of redshift information and thus without
the additional complications introduced by redshift space distortions.
The APM points are from the likelihood 
analysis of Efstathiou \& Moody (2001), using a few million galaxies,
and reflect the full uncertainty 
even on the largest scales. Here the vertical bands have a different interpretation,
indicating the bands used in the likelihood analysis. 
Note that although the 2dFGRS galaxies are a subset of the APM galaxies,
they need not have the exact same bias. Since the 2dFGRS subset involves
on average brighter and more luminous galaxies, one might expect them to 
be slightly more clustered.
The SDSS points (from Tegmark {\protect\etal} 2002) are for about a million galaxies in the magnitude range $21<r'<22$, and the vertical bars have the same interpretation
as for the 2dFGRS points (redshift information obviously helps tighten up the
windows). In contrast, the parameterized SDSS power spectrum in Dodelson {\etal} (2002) can be interpreted like the APM one.

A direct comparison of our power spectrum results with those reported by the 2dFGRS team (Percival 2001) is unfortunately not possible at this time, since
their window functions are of crucial importance and have not yet been made 
publicly available. However, an indirect comparison is possible as described
in the next section, indicating good agreement. Our $\beta$-constraints
are consistent with those reported in Peacock {\etal} (2001).

\subsection{Cosmological constraints}
\label{CosmologySec}

\begin{figure} 
\centerline{\epsfxsize=\figsize\epsffile{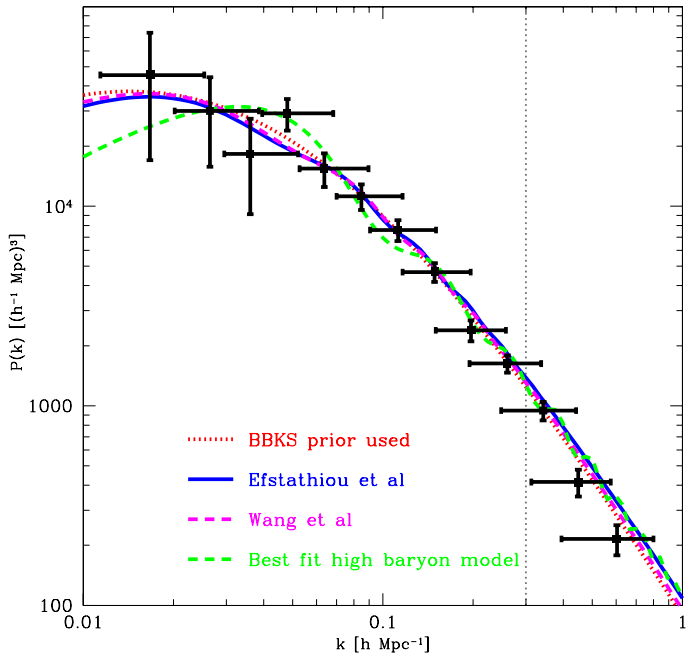}}
\caption[1]{\label{p2fig}\footnotesize%
Our 2dF power spectrum measurements from \fig{power_ggFig}
are averaged into fewer bands and compared 
with theoretical models.
The BBKS model is the wiggle-free prior used for our calculation.
The flat $\Lambda$CDM ``concordance'' models from Wang {\protect\etal} (2002) and
Efstathiou {\protect\etal} (2002), both renormalized to our 2dF measurements, 
are seen to be quite similar.
The wigglier curve corresponds to the best-fit 
high baryon model in the upper right corner of 
\fig{OmfbFig}.
Only data to the left of the dashed vertical line are included in our fits.
}
\end{figure}

\Fig{p2fig} compares our 2dFGRS measurements with theoretical 
predictions from a series of models. No corrections have been made for non-linear evolution or scale-dependent bias. 
The measurements are seen to be in good agreement
with both our simple BBKS prior and the recent concordance model
from Efstathiou {\etal} (2002) --- specifically, this is fit B from their paper, 
a flat scale-invariant scalar model with $\Ol=0.71$,
$h=0.69$, baryon density $\ob=0.021$ and dark matter density
$\od=0.12$. ($\omega_b\equiv h^2\Omega_b$,  $\omega_d\equiv h^2\Omega_d$.) 
Both of these are of course good fits by construction: 
we iterated our analysis until we found a prior that was consistent with the data, 
and Efstathiou {\etal} (2002) searched for models fitting 
both the 2dFGRS power spectrum and CMB data. 
However, the fact that the Efstathiou {\etal} (2002) model fits our data so well provides an
important cross-check between the 2dFGRS team power spectrum measurement
(Percival {\etal} 2001) and ours, indicating good agreement.

\Fig{p2fig} also shows the concordance model from Wang {\etal} (2002), resulting from
a fit to all CMB data and the PSCz galaxy power spectrum. 
It is a flat scalar model with $\Ol=0.66$,
$h=0.64$, baryon density $\ob=0.020$, dark matter density
$\od=0.12$ and a slight red-tilt, $\ns=0.91$, here renormalized to the PSCz data.
The fact that these pre-2dF and post-2dF concordance models agree so well
is a reassuring indication that such multi-parameter analyses are 
converging to the correct answer, and that the final numbers
are not overly sensitive to bias issues or methodological technicalities.

A full multiparameter analysis of our results along the lines 
of Wang {\etal} (2002) and Efstathiou {\etal} (2002) is 
clearly beyond the scope of the present paper. 
However, since evidence for baryonic wiggles in the galaxy power spectrum
has generated strong recent interest, 
first from the PSCz data (HTP00) and then more
strikingly from the 2dF data (Percival {\etal} 2001; Miller {\etal} 2001),
we perform a limited analysis to address the baryon issue. 

\begin{figure} 
\centerline{\epsfxsize=\figsize\epsffile{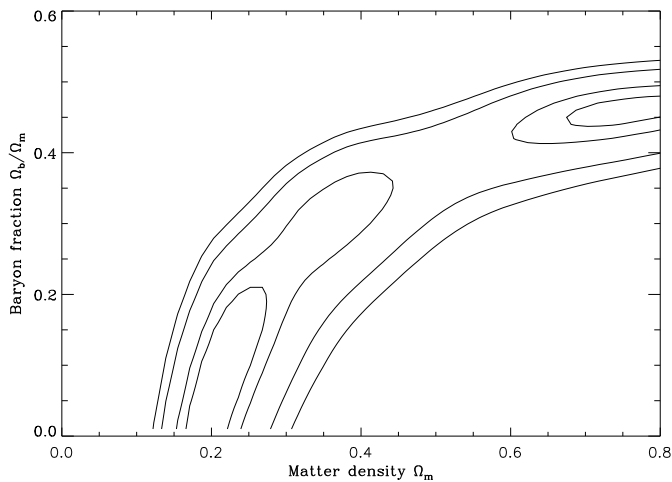}}
\caption[1]{\label{OmfbFig}\footnotesize%
Constraints in on the matter density $\Omega_m$
and the baryon fraction $\Omega_b/\Omega_m$ from
the linear power spectrum over
the range $0.01\hperMpc<k<0.3\hperMpc$, after marginalizing over
the power spectrum amplitude. 
These constraints assumes a flat, scale-invariant 
cosmological model with $h=0.72$.
For comparison with Percival et al (2001), contours
have been plotted at the level for one-parameter
confidence of 68\% and two-parameter confidence 
of 68\%, 95\% and 99\% 
(i.e., $\chi2-\chi2_{min}=1, 2.3, 6.0, 9.2$.
Marginalizing over the Hubble parameter $h$ and
limiting the analysis to scales $k<0.15h$/Mpc
as in Percival et al (2001) further weakens 
the constraints.
}
\end{figure}

We consider flat scale-invariant scalar models parametr\-ized
by the total matter content $\Om$, the baryon fraction
$\Ob/\Om$, the hubble parameter $h$ and
the spectral index $\ns$.
We map out the likelihood function $L = e^{-\chi^2/2}$ 
using \eq{chi2Eq} on a fine grid in this parameter space,
and compute constraints on individual parameters by 
marginalizing over the other parameters.
\Fig{OmfbFig} shows the result of fixing $\ns=1$ and $h=0.72$, the
best-fit value from Freedman {\etal} (2001). Here the axes have been chosen 
to facilitate comparison with Figure 5 from Percival {\etal} (2001)\footnote{
As a technical point, Percival {\etal} included band powers up to a nominal wavenumber
$k=0.15$ in their figure. Since our window functions are
narrower, we have included band powers up to  $k=0.3$ in \fig{OmfbFig} 
to ensure that we do not use less small-scale information. 
}.
The general agreement between the two figures is seen to be good, 
both in terms of the shape and location of the banana-shaped
degeneracy track, and in that there are two distinct favored regions --- a
low-baryon solution like the concordance models in \Fig{p2fig} and
a high-baryon solution that is inconsistent with both Big Bang Nucleosynthesis 
(Burles {\etal} 2001) and CMB constraints.
To illustrate the nature of the banana degeneracy in \fig{OmfbFig} , 
we have plotted the best fit high-baryon model in 
\Fig{p2fig}. It has $\Om=0.75$ and $\ob=0.18$, and is seen to
provide a slightly better fit to the data around $k=0.04\hperMpc$ at
the expense of slight difficulties on smaller scales.

There is, however, one notable difference between \fig{OmfbFig}
and its twin in Percival {\etal} (2001).
Whereas the latter 
excluded $\Omega_b/\Omega_m=0$, we find no significant detection of 
baryons. This is of course not an indication of problems with either analysis, 
since the Percival {\etal} figure excludes zero baryons only at
modest significance. Most importantly, as emphasized by Efstathiou {\etal} (2002), 
the constraints get much weaker when allowing small variations
in other parameters, most strikingly the spectral index $\ns$.
We confirm this effect by  marginalizing over $\ns$ and $h$ with various priors.
This means that the full statistical power of the complete 2dF and SDSS 
data sets will be needed to provide unequivocal evidence for baryonic 
signatures in the galaxy distribution.

\subsection{Outlook}

We have computed the real-space power spectrum and the redshift-space distortions 
of the first $10^5$ galaxies in the 2dFGRS using 
pseudo-Karhunen-Lo\`eve eigenmodes and the stochastic bias
formalism, providing easy-to-interpret uncorrelated power measurements
with narrow and 
well-behaved window functions in the range $0.01\hperMpc < k <  1\hperMpc$. 
A battery of systematic error tests
indicate that the survey is not only impressive in size, but also unusually clean.

Galaxy redshift surveys are living up to expectations.
The striking early successes of the 2dFGRS and SDSS projects
have firmly established galaxy redshift surveys as a precision tool for constraining
cosmological models. However, it is important to bear in mind that this is
only the beginning, and that many of the most exciting cosmological 
applications of these surveys still lie ahead. 
As discussed above, detailed comparisons with grids of fast simulations 
are likely to place information extracted from redshift distortions 
on a firmer footing and allow substantially more velocity information 
to be extracted from translinear scales.
A bivariate analysis of how clustering depends jointly on 
both spectral type and luminosity should improve our quantitative understanding of biasing and allow possibilities such as the above-mentioned artificial red-tilt
to be quantified and eliminated.
With such progress combined with an order-of-magnitude increase in 
sample size, to more than $10^6$ galaxies from 2dFGRS and SDSS combined,
exciting opportunities will abound over the next few years, from 
definitive constraints on baryons and neutrinos to things that have not even 
been thought of yet.

\section*{Acknowledgements}

The authors wish to thank the 2dFGRS team for kindly making the data
from this superb survey public and Shaun Cole, Mathew Colless and 
Karl Glazebrook in particular for helpful information about
technical survey details.  Thanks to Martin Kunz and Michael Vogeley
for helpful comments.
Support for this work was provided by 
NSF grants AST-0071213 \& AST-0134999,
NASA grants NAG5-9194, NAG5-11099 \& NAG5-10763,
the University of Pennsylvania Research Foundation,
the Zaccheus Daniel Foundation and the David and Lucile Packard Foundation.

\end{document}